\documentclass[iop]{emulateapj}
\usepackage{amsmath}
\usepackage{graphicx}
\usepackage{upgreek} 

\usepackage{color, xcolor}
\usepackage[colorlinks=true, linkcolor={magenta},citecolor={blue!50!black},urlcolor={magenta}]{hyperref}
\usepackage{txfonts} 
\usepackage{empheq}

\usepackage{graphicx}

\def\vecbf#1{\mbox{\boldmath $#1$}}

\newcommand{\ipa}{I_{||}}

\newcommand{\vth}{v_{\rm th}}

\newcommand{\tgas}{t_{\rm gas}}
\newcommand{\tl}{t_{\rm L}}

\newcommand{\tradp}{t_{\rm rad,\ p}}
\newcommand{\tradalign}{t_{\rm rad,\ align}}
\newcommand{\tint}{t_{\rm int}}
\newcommand{\tbar}{t_{\rm Bar}}
\newcommand{\tsup}{t_{\rm sup}}
\newcommand{\tnuc}{t_{\rm nuc}}
\newcommand{\tine}{t_{\rm ine}}

\newcommand{\aeff}{a_{\rm eff}}
\newcommand{\urad}{u_{\rm rad}}
\newcommand{\lmdbar}{\overline\lambda}

\newcommand{\kb}{k_{\rm B}}

\newcommand{\rhog}{\rho_{\rm g}}

\newcommand{\rhos}{\rho_{\rm s}}

\newcommand{\uisrf}{u_{\rm ISRF}}

\slugcomment{Accepted for publication in ApJ}

\shorttitle{Radiative grain alignment in protoplanetary disks}
\shortauthors{Tazaki et al.}

\begin{document}

\title{Radiative grain alignment in protoplanetary disks:\\ Implications for polarimetric observations}
\author{Ryo Tazaki\altaffilmark{1,2}, Alexandre Lazarian \altaffilmark{3}, and Hideko Nomura\altaffilmark{2}}
\altaffiltext{1}{Department of Astronomy, Graduate School of Science, Kyoto University, Kitashirakawa-Oiwake-cho, Sakyo-ku, Kyoto 606-8502, Japan; \email{rtazaki@kusastro.kyoto-u.ac.jp}}
\altaffiltext{2}{Department of Earth and Planetary Sciences, Tokyo Institute of Technology, 2-12-1 Ookayama, Meguro-ku, Tokyo 152-8551, Japan}
\altaffiltext{3}{Department of Astronomy, University of Wisconsin, Madison, WI 53706, USA}

\begin{abstract}
We apply the theory of radiative torque (RAT) alignment for studying protoplanetary disks around a T-Tauri star and perform 3D radiative transfer calculations to provide the expected maps of polarized radiation to be compared with observations, such as with ALMA.  We revisit the issue of grain alignment for large grains expected in the protoplanetary disks and find that mm-sized grains at midplane do not align with magnetic field as the Larmor precession timescale for such large grains becomes longer than the gaseous damping timescale. Hence, for these grains the RAT theory predicts that the alignment axis is determined by the grain precession with respect to the radiative flux. As a result, we expect that the polarization will be in the azimuthal direction for a face-on disk. It is also shown that if dust grains have superparamagnetic inclusions, magnetic field alignment is possible for (sub-)micron grains at the surface layer of disks, and this can be tested by mid-infrared polarimetric observations. 
\end{abstract}

\keywords{dust, extinction --- polarization --- protoplanetary disks --- radiative transfer}

\section{Introduction}
Magnetic fields play an important role in accretion disks at different stages. In particular, at the initial stages of disk formation, they have been proposed as a cause of the ``magnetic breaking catastrophe" that can prevent the disks from forming.
Different ideas about how to resolve this problem have so far been considered \citep[e.g.,][]{Li:2014aa}, and the most promising scenario is the reconnection diffusion  \citep[][]{Lazarian:2005aa, Lazarian:2014aa} driving the magnetic fields out \citep[see][]{Gonzalez-Casanova:2016aa}. At later stages the magneto-rotational instability \citep{Velikhov:1959aa, Chandrasekhar:1960aa, Balbus:1991aa} can play an important role in the driving the accretion. The presence of turbulence affects not only gas dynamics, but also grain dynamics. In the disk, micron-sized dust grains coagulate to form cm-sized dust aggregates, and this finally gives rise to the formation of planetesimals.
However, high-speed collisions between grains excited by disk turbulence often result in fragmentation; thereby, turbulence may halt the planetesimal formation. For these reasons, it is important to constrain the magnetic field strength and structure by observations.

Mid-infrared and millimeter polarimetric observations has so far been considered as the best method to probe the magnetic field.
This is because if aspherical grains in disks become aligned with the magnetic field as is the case in the interstellar medium (ISM), the polarization vector arising from thermal emission of the aligned grains becomes perpendicular to the local magnetic field line \citep[Cho \& Lazarian 2007, henceforth][]{CL07, Yang2016b, Matsakos:2016aa, Bertrang:2017aa}.
At mid-infrared wavelengths, \citet{Li:2016aa} performed a polarimetric imaging observation of AB Aur using CanariCam. As a result, they detected a centrosymmetric polarization pattern, and the degree of polarization was as high as 1.5\% at large radii. 
At millimeter wavelengths, polarimetric observations of disks have been performed so far \citep[e.g.,][]{Hughes:2009aa, Hughes:2013aa, Cox:2015aa, Rao:2014aa, Stephens:2014aa, kataoka2016b}. Polarized emission from a circumstellar disk has been detected in the Class 0 phase \citep{Cox:2015aa, Rao:2014aa}. 
More evolved disks do not show a degree of linear polarization larger than $0.5$\% \citep{Hughes:2009aa, Hughes:2013aa}. It should be mentioned that \citet{Stephens:2014aa} detected polarized flux from HL Tau, which is classified as a Class I-II, with an average degree of linear polarization of 0.9\%. 
More recently, \citet{kataoka2016b} reported the first submillimeter polarization observation of a disk obtained with ALMA, and they clearly detected polarized flux from HD 142527. The polarization fraction at a peak position of polarized intensity was 3.26\%, and a maximum polarization fraction was as high as 13.9\%.
The disk reveals radial polarization vectors; however, they flips by 90$^{\circ}$ in its northeast and northwest regions. 
In addition, the detected polarization fraction is much larger than the stringent limit set by \citet{Hughes:2009aa} and \citet{Hughes:2013aa}, and further polarimetric observations by ALMA will reconcile this discrepancy.

Theoretically, the origin of mid-infrared and millimeter polarization of protoplanetary disks is still a matter of debate.
In the disk, polarized radiation is expected to arise from (i) the thermal emission of aligned aspherical grains \citep{CL07} and/or (ii) the scattering of an anisotropic radiation field by dust grains \citep{kataoka2015, Kataoka:2016aa, Yang:2016aa, Yang2016b, Pohl:2016aa}.
\citet{CL07} studied grain alignment in a protoplanetary disk and they concluded that a polarization fraction of 2-3 \% at (sub-)millimeter wavelength can be expected. 
Since the magnetic field of disks is dominated by the toroidal field, grain alignment in disks has been considered to produce a radial polarization vector for a face-on disk \citep{Yang2016b, Matsakos:2016aa, Bertrang:2017aa}. Scattering polarization can be important when the maximum grain size in disks is similar to the observing wavelength \citep{kataoka2015}, and a disk inclination also helps to produce the polarized radiation by the scattering \citep{Yang:2016aa, Yang2016b}. A scattering polarization vector for the scattering basically traces the radiative flux, i.e., for a smooth face-on disk, a polarization vector will be in the azimuthal direction.

In this paper, we revisit grain alignment in disks since the work of \citet{CL07} was published before a theory of radiative torque (RAT) alignment had been formulated, i.e. before the RAT alignment paper of Lazarian \& Hoang (2007,  henceforth \citet{LH07}). Taking into account present-day RAT alignment theory \citep[see the review by][]{Lazarian:2015aa} may largely impact on their results.
Major updates of the RAT alignment theory after \citet{CL07} are as follows. Firstly, \citet{CL07} adopted an alignment condition based on \citet{DW96}; however, \citet{LH07} showed that their condition does not express the onset of grain alignment correctly. Secondly, \citet{CL07} assumed dust grains always become aligned with the magnetic field when the alignment condition is satisfied.
However,  mm-sized grains are more likely to be aligned with the radiation direction rather than with the magnetic field due to their relatively slow Larmor precession in the magnetic field \citep[][Lazarian \& Hoang in preparation]{LH07}. Thirdly, \citet{CL07} assumed perfect internal alignment, while all mechanisms of internal alignment become inefficient for sufficiently large grains \citep{Purcell79, LD99a, Lazarian:1999dy, Lazarian:2008aa}. In the absence of internal alignment, grain alignment is still possible; however, the degree of alignment, and hence the degree of polarization, is reduced \citep{HL09a}.

This paper is organized as follows.
In Section \ref{sec:alignment}, we describe our model of radiative grain alignment.
In Section \ref{sec:diskmodel}, a disk model and dust models we use in the calculations are summarized.
In Section \ref{sec:result1}, the proper alignment axis in the disk is presented based on the radiative transfer calculations and a timescale argument.
In Section \ref{sec:result2}, we estimate the degree of polarization from the disk.
In Section \ref{sec:discussion} we discuss implications for the observations, and our conclusions are presented in Section \ref{sec:summary}.

\section{Model of grain alignment} \label{sec:alignment}
In this section, we state our model for grain alignment.
In Section \ref{sec:overview}, we summarize the RAT alignment process.
In Section \ref{sec:timescale}, we describe the essential timescales for RAT alignment.
In Section \ref{sec:condition}, based on these timescales given in Section \ref{sec:timescale}, we summarize the alignment conditions.

\subsection{Overview of RAT alignment} \label{sec:overview}
\begin{figure}[tb]
\begin{center}
\includegraphics[height=6.0cm,keepaspectratio]{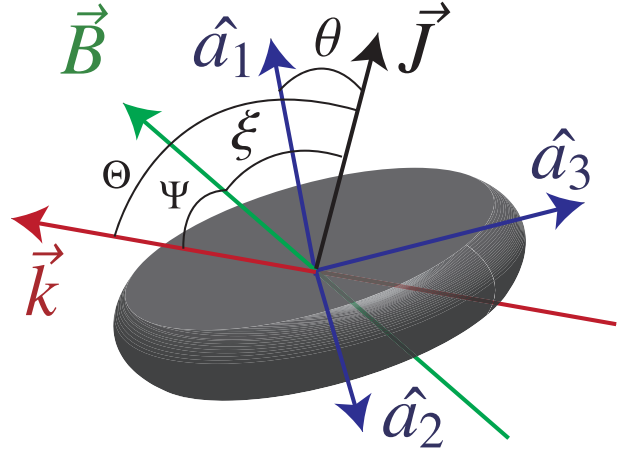}
\caption{Definition of the coordinate system. $\hat{\vecbf{a_1}}$ is the short axis of the grain, and $\hat{\vecbf{a_2}}$ and $\hat{\vecbf{a_3}}$ corresponds to the long axis of the grain. $\vecbf{J}$ is the angular momentum vector of the grain. $\vecbf{B}$ is the local magnetic field and $\vecbf{k}$ is the direction of the anisotropic radiation field, or simply the direction of the radiative flux.}
\label{fig:axis}
\end{center}
\end{figure}
In general, grain alignment involves two different alignment processes (see also Figure \ref{fig:axis}): (1) alignment of the angular momentum vector of the grains with the magnetic field ($\xi \to 0$) or alignment of the angular momentum vector of grains with the direction of the anisotropic radiation ($\Theta\to 0$), and (2) alignment of the minor axis of the grains with the angular momentum vector ($\theta\to 0$). These alignment processes are referred to as {\it external alignment} and {\it internal alignment}. 

The driving physics of external alignment is RAT.  
RAT alignment naturally explains the observed features of the interstellar polarization \citep[see the review by][]{Andersson2015}.
The presence of RAT was firstly realized by \citet{dolginov76} and confirmed by numerical simulations by \citet{DW96, DW97}.  
\citet{DW96} found that RAT under the isotropic radiation can spin up a grain suprathermally, and the subsequent paramagnetic dissipation \citep{DG51} leads to grain alignment with respect to the magnetic field. However, RAT arising from the isotropic radiation is fixed in grain coordinates and this prevents grains from rotating suprathermally due to the effect known as thermal flipping \citep{LD99b, HL09b}.
\citet{DW97} argued that dust grains can be spun up to suprathermal under an anisotropic radiation field. In addition, they also found that in this case, RAT can drive external alignment even in the absence of paramagnetic dissipation. 
However, calculations taking into account a more accurate treatment of crossover 
\footnote{Disregard of the crossover physics resulted in the appearance of cyclic phase trajectories in \citet{DW97} and \citet{WD03}. This cyclic behavior is not physical. Instead, the grains enter the low-J attractor point \citep{WD03, LH07}.} 
showed that RAT under an anisotropic radiation field often spin down the grain rotation to thermal or subthermal \citep{WD03, LH07}. 
An important property of RAT alignment is that a system of dust grains evolves into stable points in the parameter space of the angular momentum and the alignment angle. These stable points are often referred to as the ``attractors" \citep{DW97}. They often (but not always) appear at perfect alignment i.e., $\xi=0$ and $\Theta=0$ \citep{DW97,LH07}. 
There exist two kinds of attractors: one is high-$J$ attractors (a spin-up state) whose angular momentum is suprathermal, and the low-$J$ attractors (a spin-down state) whose angular momentum is thermal or subthermal. High-$J$ and low-$J$ attractors can appear simultaneously in the phase trajectory map; however, even in this case, only a small fraction of grains reaches to the high-$J$ attractor. 
For this reason, it turns out that suprathermal rotation driven by RAT should not be used as a necessary condition for the alignment \citep[see also the review by][]{Lazarian07}. It is worth noting that the appearance of high-$J$ or low-$J$ attractors can be predicted by the $q^{\rm max}$-parameter and $\Psi$ (see Figure 24 of \citet{LH07}), where the $q^{\rm max}$-parameter describes the grain morphology, or the grain helicity. 

A necessary condition of RAT alignment is the precession motion of a grain. This is essential for grain alignment because the precession axis defines the alignment axis regardless of the origin of precession. If dust grains are forced to quit precession due, for example, to the gaseous damping, alignment does not occur. This is because the grain precession motion can stabilize the alignment torque.
The major origin of grain precession is Larmor precession or radiative precession (see Section \ref{sec:timescale} for more details).
If Larmor precession governs the grain precession motion, dust grains become aligned with the magnetic field, as is the case in the ISM. On the other hand, if radiative precession overcomes Larmor precession, grains become aligned with the direction of radiation, not the magnetic field \citep{LH07}.
The above basic physics should be augmented by additional important points. First of all, the grain alignment at the high-$J$ attractor point can be stabilized by enhanced magnetic susceptibility arising from superparamagnetic or ferromagnetic inclusions, e.g. iron clusters \citep{Lazarian:2008aa, HL16}. The latter alignment is perfect, unlike the result for low-$J$ alignment, which does not usually exceed 30\% \citep{HL08}. In addition, suprathermal torques introduced originally by \citet{Purcell79}, if present in circumstellar disks, can also increase the degree of RAT alignment \citep{HL09b}.

Internal alignment, or internal relaxation, is a process where the grain axis for the maximum inertia aligns with the angular momentum vector \citep{Purcell79}. 
The rotational kinetic energy of a body can be given by \citep[see][]{Lazarian:1997aa}
\begin{equation}
E(\theta)=\frac{J^2}{I_{\rm ||}}(1+\sin^2\theta(h-1)),
\end{equation}
where $h=I_{||}/I_{\perp}$, $I_{||}$ and $I_{\perp}$ represent the inertia with respect to the grain minor axis and major axis, $J$ is the angular momentum of the grain, and $\theta$ is the angle between $\hat{\vecbf{a_1}}$ and $\vecbf{J}$.
The rotational kinetic energy of a body can be converted into heat by internal energy dissipation.
Since the angular momentum of a body $J$ is conserved when we consider torque-free motion of a rigid body, energy dissipation leads to alignment of the axis of the greatest inertia of the grains with its angular momentum vector. Various energy dissipation processes have so far been proposed: Barnett relaxation \citep{Purcell79}, nuclear relaxation \citep{LD99a}, superparamagnetic Barnett relaxation \citep{LD99a, HL08}, and inelastic relaxation \citep{Purcell79, Lazarian:1999dy}. 

\subsection{Timescales relevant to RAT alignment} \label{sec:timescale}
In this study, we use the timescales to specify the degree of grain alignment instead of solving the equation of motion of an aspherical grain under an anisotropic radiation field. We summarize the relevant timescales of external alignment and internal alignment.
\subsubsection{Grain geometry}
Modeling of the interstellar polarization has been suggested that an oblate grain is preferred to a prolate grain \citep{Lee:1985aa, Henning:1993aa, Kim:1995aa, Hildebrand:1995aa}.
In addition, \citet{Hildebrand:1995aa} showed that an oblate with axis ratio of 2:3 produces the best fit for far-infrared polarization.
Although dust grains in disks may differ from interstellar grains, in this paper, we adopt an oblate spheroid for simplicity.
Denoting the inertia with respect to the grain minor axis and major axis by $I_{||}$ and $I_{\perp}$, respectively, then
\begin{equation}
I_{||}=\frac{8\pi}{15}\rho_{\rm s}a_1a_2^4,\ I_{\perp}=\frac{4\pi}{15}\rho_{\rm s}a_1a_2^2(a_1^2+a_2^2),
\end{equation}
where $\rho_s$ is the material density, $a_1,\ a_2$ denotes the minor and major radii of a grain, and then the grain aspect ratio $s$ is defined by $s=a_1/a_2$.
It is useful to define a characteristic radius of an aspherical grain. In this paper, we use the volume-equivalent radius $\aeff$ for which the volume of sphere with radius $\aeff$ equals to that of the original ellipsoid; hence, $\aeff^3\equiv a_1a_2^2$.

\subsubsection{Gaseous damping timescale} \label{sec:gaseous}
Random collisions of gas particles prevents grains from being aligned.
\citet{Roberge:1993aa} derived the gaseous damping timescale assuming perfect sticking of colliding molecules and subsequent evaporation of molecules from the surface, which are assumed to be thermalized at the grain surface.
Although the mean torque due to sticking collision is canceled out, that due to evaporation is non-zero for spinning grains. Thus, evaporated molecules extract angular momentum from spinning dust grains.
The timescale for which grain angular momentum becomes zero via the gaseous damping is given by
\begin{eqnarray}
\tgas&=&\frac{3}{4\sqrt{\pi}}\frac{\ipa}{\rhog \vth a_2^4\Gamma_{||}} \label{eq:tgas}\\
&\approx& 8.1 \times10^{-3}\ \hat{\rhos}a_{-5}\hat{\rhog}^{-1}\hat{T_g}^{-1/2}\ {\rm yr}
\end{eqnarray}
where $\rhog$ is the gas density, $T_g$ is the gas temperature, $\rhos$ is the material density, $\vth=\sqrt{2k_{\rm B}T_{\rm g}/\mu m_{\rm H}}$, where $\mu=2.34$ is a mean molecular weight.  $\Gamma_{||}$ is a geometrical coefficient given by
\begin{eqnarray}
 \Gamma_{||}&=&\frac{3}{16}\{3+4(1-e_m^2)\tilde{g}(e_m)-e_m^{-2}[1-(1-e_m^2)^2\tilde{g}(e_m)]\}\nonumber \\
&& \\
 \tilde{g}(e_m)&=&\frac{1}{2e_m}\ln\left(\frac{1+e_m}{1-e_m}\right)
\end{eqnarray}
where $e_m^2=1-s^2$. We adopt the following normalization: $\hat{\rhos}=\rhos/3\ {\rm g\ cm}^{-3}$, $\hat{s}=s/0.5$, $a_{-5}=\aeff/10^{-5}\ {\rm cm}$, $\hat{\rhog}=\rhog/10^{-15}$ g cm$^{-3}$, $\hat{T_g}=T_g/100$ K. Note that we disregard infrared emission damping \citep{Purcell:1971aa} and the plasma drag \citep{Draine:1998aa}, because this mechanism only becomes important for small grains whose size is less than $0.1\mu$m, while we expect more larger grains in disks.

\subsubsection{Precession of $\vecbf{J}$ around $\vecbf{B}$: Larmor precession timescale}
A magnetized body precesses around an applied magnetic field, and this is called Larmor precession.
Even in the absence of a spontaneous magnetic moment, spinning dust grains can acquire a magnetic moment via (i) surface charges \citep{Martin:1971aa} and (ii) the Barnett effect \citep{dolginov76}. If a dust grain has non-zero surface charges, the spinning grain generates a magnetic dipole moment via the surface current, and hence the grain becomes magnetized. The Barnett effect \footnote{The Barnett effect is a reciprocal phenomenon of the Einstein--de Haas effect in which magnetization of a body induces mechanical rotation \citep[see, e.g.,][]{Landau:1960aa}.} is a phenomenon in which a rotating body becomes magnetized with the magnetic moment parallel to the angular velocity \citep{Barnett:1915aa}. In most cases, a magnetic moment induced by grain rotation is dominated by the Barnett effect. Thus, in this paper, we neglect the magnetization by surface charges.

The Larmor precession timescale, or the timescale of precession of the magnetic moment induced by grain rotation around the magnetic field $B$ can be given by
\begin{equation}
t_{\rm L}=\frac{2\pi}{\Omega_{\rm L}},\ \Omega_{\rm L}=\frac{\mu_{\rm Bar}B}{I_{||}\omega} \label{eq:tl}
\end{equation}
where 
$\Omega_{\rm L}$ is the angular frequency of Larmor precession, $\omega$ is the angular frequency of a grain rotation, and $\mu_{\rm Bar}$ is the magnetic moment induced by the Barnett effect \citep{Landau:1960aa}:
\begin{equation}
\mu_{\rm Bar}=\frac{\chi(0)V\hbar}{g\mu_{\rm B}}\omega \label{eq:mubar}
\end{equation}
where $V$ is the volume of the grain, $\hbar$ is the Planck constant divided by $2\pi$, and $\mu_{\rm B}=e\hbar/2m_ec\approx 9.274\times 10^{-21}$ erg G$^{-1}$ is the Bohr magneton, where $m_e$ is the mass of the electron, and $c$ is the speed of light, and $g$ is the $g$-factor, which is $g\approx 2$ for electrons. 
$\chi(0)$ is the magnetic susceptibility at zero frequency.
From Equations (\ref{eq:tl}) and (\ref{eq:mubar}), we obtain
\begin{eqnarray}
t_{\rm L}&=&\frac{4\pi}{5} \frac{g_e\mu_{\rm B}}{\hbar}\rho_s s^{-2/3}a_{\rm eff}^{2}B^{-1}\chi(0)^{-1}\\
&\approx& 1.3\ \hat{\rhos}\hat{s}^{-2/3}a_{-5}^2\hat{B}^{-1}\hat{\chi}^{-1}\ {\rm yr}
\end{eqnarray}
where $\hat{B}=B/5\ \mu$G, $\hat{\chi}=\chi(0)/10^{-4}$.

The magnetic susceptibility of dust grains depends on the grain's material properties.
Most materials have unpaired electrons in the outer, partly filled shell, and these electrons can contribute to the grain magnetic moment.
If the interaction between spins can be ignored, the material is regarded as paramagnetic. 
A property of paramagnetic grains is that in the absence of an applied magnetic field, the number of up-spins and down-spins is statistically the same; hence, a paramagnetic material does not have a spontaneous magnetic moment. However, once the field is applied, it becomes magnetized. Suppose $f_p$ is the fraction of atoms in the grain that are paramagnetic; then the zero-frequency susceptibility is given by the Curie's law \citep{Morrish2001},
\begin{equation}
\chi(0)=\chi_e(0)=4.2\times10^{-2} f_p\left(\frac{T_d}{15\ {\rm K}}\right)^{-1} \label{eq:chie}
\end{equation}
As the grain temperature increases, the spin thermal fluctuation of spins halts the magnetization, and then the magnetic susceptibility is reduced. Observations of interstellar depletion suggest that $10$\% of grain atoms are Fe atoms; then $f_p$ is as large as 0.1 \citep[e.g.,][]{Draine1996}.
\citet{JS67} proposed that the magnetic susceptibility of grains can be largely enhanced by the presence of superparamagnetic inclusions.
An example of a superparamagnetic inclusion is a nanocluster of iron atoms. 
Iron atoms in a cluster interact with neighboring atoms by an exchange interaction such as ferromagnetism, and then a single cluster behaves as if it were a single atom having a large magnetic moment.
As a result, this grain behaves as if it were ferromagnetic under the presence of an applied field.
The volume of the cluster is small enough such that its magnetic moment of a single cluster readily flips due to thermal fluctuation; therefore, a superparamagnetic grain does not have a spontaneous magnetic moment.
Hence, dust grains with superparamagnetic inclusions behaves like paramagnetic grains in the absence of an applied field, and ferromagnetic grains in the presence of the field. The zero-frequency susceptibility for a superparamagnetic grain is given by the Curie's law \citep{Morrish2001},
\begin{equation}
\chi(0)=\chi_{\rm sup}(0)=1.2\times10^{-2}N_{\rm cl}\phi_{\rm sp} \left(\frac{T_d}{15\ {\rm K}}\right)^{-1}, \label{eq:chisup}
\end{equation}
where $\phi_{\rm sp}$ is the fraction of atoms that are superparamagnetic.  $N_{\rm cl}$ is the number of atoms per cluster.
The typical value from measurement of GEMS suggests $\phi_{\rm sp}=0.03$ \citep{Bradley:1994aa, Martin:1995aa}, and $N_{\rm cl}$ is expected to have $N_{\rm cl}=10^{3}-10^{5}$ \citep{JS67}.

\subsubsection{Precession of $\vecbf{J}$ around $\vecbf{k}$: Radiative precession timescale}
When dust grains are immersed in an anisotropic radiation field, they experience RAT.
The role of RAT is give rise to spin up (down), alignment, and to induce grain precession. 
\citet{LH07} identified RAT alignment with respect to the radiative flux rather than the magnetic field in the situations where the grain precession rate induced by RAT exceeds that induced by Larmor precession. The RAT is defined by \citep{DW96}
\begin{equation}
\mathbf{\Gamma}_{\rm rad}=\frac{u_{\rm rad}a_{\rm eff}^2\overline{\lambda}}{2}\overline{\gamma}\overline{\mathbf{Q_{\Gamma}}}
\end{equation}
where a bar indicates the spectrum-averaged quantities, 
\begin{eqnarray}
&& \overline{\mathbf{Q_{\Gamma}}}=\frac{\int \mathbf{Q_{\Gamma}}u_{\lambda} d\lambda}{u_{\rm rad}}, \
\overline{\lambda}=\frac{\int u_{\lambda}\lambda d\lambda}{u_{\rm rad}},\\
&& \overline{\gamma}=\frac{\int u_{\lambda}\gamma_{\lambda} d\lambda}{u_{\rm rad}},\
 u_{\rm rad}={\int u_{\lambda}d\lambda}
\end{eqnarray}
where $u_{\rm \lambda}$ is the energy spectrum of the radiation field, $\lambda$ is the radiation wavelength, $\gamma_\lambda$ is the anisotropy parameter, and $\mathbf{Q_{\Gamma}}$ is the RAT efficiency. The amplitude of $\mathbf{Q_{\Gamma}}$ is estimated by the DDA calculation \citep{LH07}, 
\begin{eqnarray}
|\mathbf{Q_{\Gamma}}|&\approx& 2.3\left(\frac{\lambda}{\aeff}\right)^{-3}\ \ {\rm for\ } \lambda>1.8\aeff\\
&\approx& 0.4\ \ \ \ \ \ \ \ \ \ \ \ \ {\rm for\ } \lambda\leq1.8\aeff \label{eq:QRATs}
\end{eqnarray}

The radiative precession timescale can be written as \citep{LH07} \footnote{There is a typographical error in their Equation (85).}
\begin{eqnarray}
t_{\rm rad,\ p}&=&\frac{2\pi}{\Omega_{\rm p}}, \label{eq:tradp} \\
\Omega_{\rm p}&=& \frac{u_{\rm rad}\bar{\lambda} a_{\rm eff}^2}{I_{||}\omega}\gamma\overline{|\mathbf{Q_{\Gamma}}|}
\end{eqnarray}
where $I_{||}$ is the maximum moment of inertia, and $\omega$ is the angular velocity of a grain.

\begin{eqnarray}
\tradp&\approx&1.1\times10^{2}\ \hat{\rhos}^{1/2}\hat{s}^{-1/3}a_{-5}^{1/2}\hat{T_d}^{1/2}\\
&\times&\left(\frac{\urad}{\uisrf}\right)^{-1}\left(\frac{\lmdbar}{1.2\ \mu{\rm m}}\right)^{-1}\left(\frac{\gamma \overline{|\mathbf{Q_{\Gamma}}|}}{0.01}\right)^{-1}\ {\rm yr},\nonumber  
\end{eqnarray}
where we have assumed the thermal angular velocity $\omega=\sqrt{2\kb T_{\rm d}/\ipa}$.

\subsubsection{Internal alignment timescale}
To estimate the internal alignment timescale, we assume the torque-free motion. As we discussed in Section \ref{sec:gaseous}, gas particles exerts torque on dust grains; hence, an approximation of the torque-free motion is valid when we consider a timescale shorter than that of the gaseous damping. The timescale for internal alignment is given by \citep{Purcell79, Roberge:1993aa, Lazarian:1997aa},
\begin{equation}
t_{\rm int}=G\frac{\gamma^2}{K(\omega)},\ G=\frac{I_{||}^3}{h^2(h-1)VJ_{\rm T}^2}\left(\frac{J_{\rm T}}{J}\right)^2 \label{eq:tint}
\end{equation}
where $\gamma$ is the gyromagnetic ratio, and $J_{\rm T}=\sqrt{I_{||}k_BT_d/(h-1)}$ is the thermal angular momentum of the grains. 
$K$ is the ratio of the imaginary part of the magnetic susceptibility and frequency:
\begin{equation}
K\equiv\frac{\chi''}{\omega}=\chi(0)\frac{\tau}{[1+(\omega\tau/2)^2]^2} \label{eq:K}
\end{equation}
where $\tau$ is the timescale of the relaxation \citep{Morrish2001}.
Note that the integration of $K(\omega)$ over zero frequency to infinity gives simply give $\chi(0)$ because of the requirement of the Kramers-Kr\"onig relationship \citep[e.g.,][]{Landau:1960aa}.
As energy dissipation proceeds, the minor axis of the grains becomes aligned with its angular momentum vector. 
The following relaxation mechanisms can be considered: Barnett relaxation, nuclear relaxation, superparamagnetic Barnett relaxation, and inelastic relaxation. 
The timescale for each process can be estimated as described below.

Barnett relaxation \citep{Purcell79} is due to relaxation of the electron's spin--spin interaction.
Hence, the gyromagnetic ratio in Equation (\ref{eq:tint}) is given by the electron, then
\begin{equation}
\gamma=\gamma_e=\frac{\mu_e}{\hbar}=\frac{g_e\mu_{\rm B}}{\hbar}, \label{eq:ge}
\end{equation}
where $g_e\approx 2$, and $\mu_{\rm B}=e\hbar/2m_ec$ is the Bohr magneton.
$\tau=\tau_e=2.9\times10^{-12}f_p^{-1}$ s is the spin-spin coupling time \citep{Morrish2001}.
Using Equations (\ref{eq:tint}), (\ref{eq:K}), (\ref{eq:ge}), and (\ref{eq:chie}), we obtain the timescale for internal alignment due to Barnett relaxation as
\begin{eqnarray}
\tbar&\equiv& G\frac{\gamma_e^2}{K_e}\\
&\approx&2.3\ \hat{\rhos}^2a_{-5}^{7}\left(\frac{J_T}{J}\right)^2\left[1+\left(\frac{\omega\tau_e}{2}\right)^2\right]^2\ {\rm yr}.\label{eq:tbar}
\end{eqnarray}

\citet{LD99a} found that the nuclear--electron spin interaction and nuclear--nuclear spin interaction lead to dissipation of the energy and this gives rise to internal relaxation.
In the case of nuclear relaxation, 
\begin{equation}
\gamma=\gamma_n=\frac{\mu_n}{\hbar}=\frac{g_n\mu_{\rm N}}{\hbar},
\end{equation}
where $\mu_{\rm N}=e\hbar/2m_pc$ is the nuclear magneton and $m_p$ is the mass of the proton.
The zero-frequency susceptibility is be given by the Curie's law \citep{Morrish2001},
\begin{equation}
\chi(0)=\chi_n(0)=4.1\times10^{-11} \left(\frac{\mu_n}{\mu_N}\right)^2\left(\frac{n_n}{10^{22}\ {\rm cm}^{-3}}\right)\left(\frac{T_d}{15\ {\rm K}}\right)^{-1}.
\end{equation}
The relaxation timescale of nuclear spin is given by 
$\tau_{\rm n}^{-1}=\tau_{\rm nn}^{-1}+\tau_{\rm ne}^{-1}$, where the timescales for nuclear--electron spin and nuclear--nuclear spin interactions are described by
\begin{eqnarray}
\tau_{\rm ne}&=&3.0\times10^{-4}\left(\frac{2.7}{g_n}\right)^2\left(\frac{n_n}{10^{22}\ {\rm cm}^{-3}}\right)\\
\tau_{\rm nn}&=&0.58\tau_{\rm ne}\left(\frac{n_e}{n_n}\right),
\end{eqnarray}
respectively \citep{LD99a}. 
As a result,
\begin{eqnarray}
\tnuc&\equiv& G\frac{\gamma_n^2}{K_n}\\
&\approx&4.6\times10^{-6}\ \hat{\rhos}^2a_{-5}^{7}\left(\frac{J_T}{J}\right)^2\left[1+\left(\frac{\omega\tau_n}{2}\right)^2\right]^2\ {\rm yr} \nonumber \\
\end{eqnarray}

In the presence of superparamagnetic inclusions, superparamagnetic Barnet relaxation is important. For this process, the gyromagnetic ratio is given by $\gamma=\gamma_e$.
The relaxation timescale is
\begin{equation}
\tau_{\rm sup}^{-1}=\nu_0e^{-N_{\rm cl}\theta/T}
\end{equation}
where $\nu_0=10^9$ s$^{-1}$, and $\theta=0.011$ K \citep{Morrish2001}.
As a result,
\begin{eqnarray}
\tsup&\equiv& G\frac{\gamma_e^2}{K_{\rm sup}}\\
&\approx&4.2\times10^{-4}\ \hat{\rhos}^2a_{-5}^{7}\left(\frac{J_T}{J}\right)^2\left[1+\left(\frac{\omega\tau_{\rm sup}}{2}\right)^2\right]^2\ {\rm yr} \nonumber \\
\end{eqnarray}
where we have used $\phi_{\rm sp}=1$ \% and $N_{\rm cl}=2\times10^{3}$.

\citet{Lazarian:1999dy} argued that the deformation induces important energy dissipation.
The timescale of inelastic dissipation is
\begin{eqnarray}
\tine &\approx& 2^{11}3^{-2.5}g(s)\aeff^{11/2}\left(\frac{J}{J_T}\right)^{-3/2}(\kb T_g)^{-3/2}\mu \mathcal{Q} \rhos^{1/2} \nonumber \\
\\
g(s)&=&\frac{s^{-1/3}[1+s^2]^4}{64s^4+20}
\end{eqnarray}
where $\mathcal{Q}$ is the inelastic parameter, and $\mu$ is rigid module.
For a silicate grain with axis ratio 1:2 ($s=0.5$), we obtain
\begin{equation}
\tine\approx 8.7 a_{-5}^{11/2}\left(\frac{J}{J_T}\right)^{-3/2}\hat{T_d}^{-3/2}\hat{\mu}\hat{\mathcal{Q}}\hat{\rhos}^{1/2}\ {\rm yr}
\end{equation}
where $\hat{\mathcal{Q}}=\mathcal{Q}/100$, $\hat{\mu}=\mu/10\ {\rm Pa}$.

Internal alignment is caused by the complex of energy dissipation processes described above; then the internal alignment timescale is
\begin{equation}
\frac{1}{\tint} \approx \frac{1}{\tbar}+\frac{1}{\tnuc}+\frac{1}{\tsup}+\frac{1}{\tine}
\end{equation}

\begin{figure}[tb]
\begin{center}
\includegraphics[height=7.0cm,keepaspectratio]{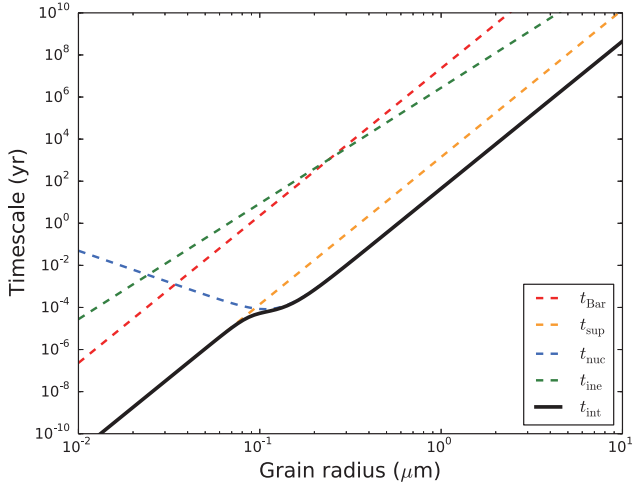}
\caption{Timescale of internal alignment. Adopted parameters are $s=0.5$, $J=J_{T}$, $f_p=0.1$, $\phi_{\rm sp}=0.01$, and $N_{\rm cl}=2\times10^3$.}
\label{fig:internal}
\end{center}
\end{figure}
Figure \ref{fig:internal} shows the internal relaxation timescale as a function of the grain radius. In the presence of superparamagnetic inclusions, internal relaxation is dominated by superparamagnetic Barnett relaxation for smaller grains and by nuclear relaxation for larger grains. Since $t_{\rm nuc}\propto \aeff^7$, internal relaxation does not likely to occur for larger grains, e.g., mm-size grains for which the timescale is longer than the age of the universe. 

\subsection{Conditions for the alignment} \label{sec:condition}
\citet{CL07} assumed alignment happens when the energy density of the radiation field is large enough to spin up grains to a suprathermal state according to \citet{DW96}. However, as we mentioned in Section \ref{sec:overview}, the suprathermal rotation is not always necessary for grain alignment.
In addition, \citet{CL07} assumed the alignment axis is always parallel to the local magnetic field, but this is not true because grains can be aligned with the radiation direction \citep{LH07}.
Here, we state the alignment condition adopted in this paper.

\subsubsection{Alignment axis}
The precession motion of a grain is essential for grain alignment because the precession axis will be the alignment axis.
Denoting the timescale of precession $t_{\rm p}$, which is adopted as a minimum value of the Larmor precession, $t_{\rm L}$ (Equation \ref{eq:tl}) or the radiative precession, $t_{\rm rad,p}$ (Equation \ref{eq:tradp}), we have
\begin{equation}
t_{\rm p}={\rm min}(t_{\rm rad,p},t_{\rm L})
\end{equation}
The grain alignment occurs when $t_{\rm p}<t_{\rm gas}$, where $t_{\rm gas}$ is the gaseous damping timescale. 
If $t_{\rm gas}<t_{\rm p}$, the precession motion of a grain is suppressed, hence no alignment is possible.
As shown by \citet{LH07}, when Larmor precession dominates radiative precession, or $\tradp/\tl>1$, the grain angular momentum vector $\vecbf{J}$ is aligned with respect to the direction of the local magnetic field $\vecbf{B}$ ($\xi\to0$ in Figure \ref{fig:axis}). When radiative precession dominates Larmor precession, or $\tradp/\tl<1$, the grain angular momentum vector $\vecbf{J}$ is aligned with the direction of the anisotropic radiation $\vecbf{k}$ ($\Theta \to 0$ in Figure \ref{fig:axis}). 

The timescale of fast alignment \footnote{\citet{LH07} regarded fast alignment as an analogy of a fast dynamo, since RAT alignment can take place faster than the gaseous damping timescale.} to $\vecbf{B}$ or $\vecbf{k}$ due to RAT is somewhat longer than the precessions timescale \citep[see Figure 29 in][]{LH07}, and 
\begin{equation}
t_{\rm rad,\ align}\approx Ct_{\rm p} \label{eq:tradalign}
\end{equation}
where $C=30 \sim 100$. In this work, we adopt $C=30$.
Equation (\ref{eq:tradalign}) expresses the timescale of grain alignment with respect to both the magnetic field and the radiative flux.

\subsubsection{Degree of alignment}
From a statistical point of view, grain alignment means that the distribution of an alignment angle $\eta$, which is the angle between the grain axis $\hat{a_1}$ and the precession axis (the magnetic field vector $\vec{B}$ or the anisotropic radiation $\vec{k}$), should be $\langle \cos^2\eta\rangle < 1/3$ where the brakets denote the average over the ensemble of dust grains. For randomly orientated grains, $\langle \cos^2\eta\rangle=1/3$.
We define the degree of RAT alignment by $Q_{\rm RATs}=(3\langle \cos^2\eta\rangle-1)/2$. In the same way, we define the degree of internal alignment by $Q_{\rm int}=(3\langle \cos^2\theta\rangle-1)/2$. 

We write the degree of grain alignment $\mathcal{R}(a)$ of distribution of dust grains approximately as
\begin{equation}
\mathcal{R}(a)\approx f_{{\rm high-}J}Q_{{\rm RATs}}^{{\rm high-}J}Q_{\rm int}^{{\rm high-}J}+(1- f_{{\rm high-}J})Q_{\rm RATs}^{{\rm low-}J}Q_{\rm int}^{{\rm low-}J} \label{eq:doa}
\end{equation}
where $f_{{\rm high-}J}$ is the fraction of grains in high-$J$ attractors. 
$f_{{\rm high-}J}$ depends on the grain size, shape, and optical properties.
Since the properties of dust aggregates in protoplanetary disks are poorly understood, we treat $f_{{\rm high-}J}$ as a free parameter throughout this paper.
Hence, $f_{{\rm high-}J}=1$ corresponds to the case for grains to be aligned most efficiently (the optimistic case), while $f_{{\rm high-}J}=0$ is the most difficult condition for grains to be aligned (the pessimistic case). 

At high-$J$ attractors, the degree of RAT alignment does not depend on internal relaxation \citep{LH07,HL09b}; hence we can approximately write
\begin{equation}
Q_{{\rm RATs}}^{{\rm high-}J} = \left\{ \begin{array}{ll}
 1, 
 & (t_{\rm rad,align}< t_{\rm gas}) \\[3mm]
 0,
 & (t_{\rm rad,align}\geq t_{\rm gas}).
\end{array} \right.
\label{eq:alignhighJ}
\end{equation}
On the other hand, at low-$J$ attractors, the degree of RAT alignment does depend on internal alignment \citep{HL09a}.
Then, we adopt
\begin{equation}
Q_{{\rm RATs}}^{{\rm low-}J} = \left\{ \begin{array}{ll}
 0.2, 
 & (t_{\rm int}<t_{\rm rad,align}<t_{\rm gas}) \\[3mm]
 0.1, 
 & (t_{\rm rad,align}< t_{\rm int}<t_{\rm gas}) \\[3mm]
 0.1, 
 & (t_{\rm rad,align}< t_{\rm gas}<t_{\rm int}) \\[3mm]
 0,
 & (t_{\rm rad,align}\geq t_{\rm gas}), 
\end{array} \right.
\label{eq:p}
\end{equation}

Since perfect internal alignment minimizes their rotational kinetic energy, suprathermally rotating grains readily become internally aligned through transferring their rotational kinetic energy into vibrational energy \citep[e.g.,][]{Draine2011}. 
In addition, grains in a state of high-$J$ attractors are not likely to change their angular momentum vector by gas collisions because each of gas particle deposits only a small amount of angular momentum compared to the grain angular momentum.
Hence, the degree of internal alignment at high-$J$ attractors is always expected to be almost perfect; then we set $Q_{{\rm int}}^{{\rm high-}J} = 1$.
On the other hand, at low-$J$ attractors, the rotational kinetic energy can be affected by its own thermal fluctuation and also gas bombardment. Therefore, in this case, the degree of internal alignment should not be 100\%.
In the absence of internal alignment, low-$J$ attractors behave in two different ways: some attractors are aligned parallel to $\vecbf{J}$ (``right" alignment), while the others are aligned perpendicular to $\vecbf{J}$ (``wrong" alignment) \citep{HL08}.
We assume $Q_{{\rm int}}^{{\rm low-}J} = 0.5$. A more detailed calculation of $Q_{\rm int}$ for dust aggregates in the absence of internal relaxation is necessary, but this is beyond the scope of this study.
Table \ref{tab:doa} shows the degree of alignment $\mathcal{R}(a)$ calculated with Equation (\ref{eq:doa}).

\begin{table}
\caption{Model of degree of alignment, $\mathcal{R}(a)$, defined by Equation (\ref{eq:doa})}
\label{tab:r}
\centering
\begin{tabular}{llcllcllcllcll}
\hline
${\rm timescales}$ & $f_{{\rm high-}J}=1$ && $f_{{\rm high-}J}=0.5$ & $f_{{\rm high-}J}=0$ \\
\hline  \hline
$t_{\rm gas}<t_{\rm p}$ &$0$ \%&&$0$ \%&$0$ \%\\
$t_{\rm int}<t_{\rm rad,align}<t_{\rm gas}$ &$100$ \%&&$55$ \%&$10$ \%\\
$t_{\rm rad,align}<t_{\rm int}<t_{\rm gas}$ &$100$ \%&&$52.5$ \%&$5$ \%\\
$t_{\rm rad,align}<t_{\rm gas}<t_{\rm int}$ &$100$ \%&&$50$ \%&$0$ \%\\
\hline
\end{tabular}
\label{tab:doa}
\end{table}

\section{Disk and Dust models} \label{sec:diskmodel}
\subsection{The disk model and method}
The surface density profile of dust grains in the disk is assumed to be a similarity solution, given by
\begin{equation}
\Sigma_d=(2-\zeta)\frac{M_d}{2\pi R_c^2}\left(\frac{R}{R_c}\right)^{-\zeta}\exp\left[-\left(\frac{R}{R_c}\right)^{2-\zeta}\right]
\end{equation}
where $M_d$ is the disk dust mass, $R_c$ is the cut-off radius, and $R$ is the disk radius \citep{LBP74}. For simplicity, the vertical distribution of dust grains is assumed to be Gaussian, $\rho_d=\Sigma_d/(\sqrt{2\pi}H_d)\exp[-(z^2/2H_d^2)]$, where $H_d$ is the dust scale height.
The parameters adopted in this study are as follows: $M_{d}=10^{-4}\ M_{\odot}$, $\zeta=1.0$, $R_{\rm in}=0.1$ au, $R_{\rm out}=100$ au, $R_c=20$ au, $H_d=3.3\times10^{-2} (R/1\ {\rm au})^{1.25}$ au.
The central star is assumed to be a T-Tauri star with a mass of $0.5 M_{\odot}$, the effective temperature of $4000$ K, and the radius of $2R_{\odot}$.
Note that although the mass of the star does not come into the radiative transfer calculations, it is used in the dust density calculation through the Kepler frequency in Section \ref{sec:diskpolarization}.

\begin{figure}[tb]
\begin{center}
\includegraphics[height=6.0cm,keepaspectratio]{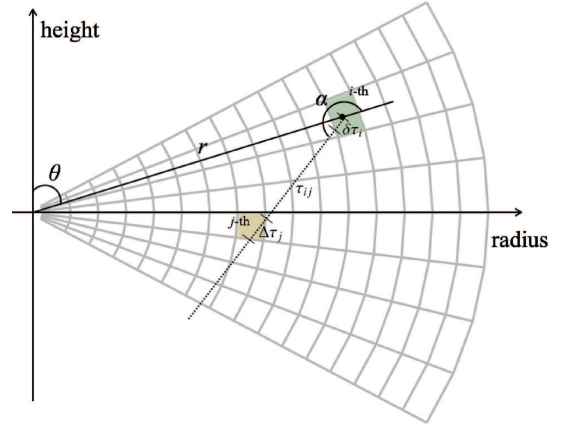}
\caption{Schematic illustration of integration of intensity along the line of sight. We use spherical coordinates with a $N_{r}=256$, $N_{\theta}=128$ mesh.}
\label{fig:grid}
\end{center}
\end{figure}

Based on the star and disk model, we perform a radiative transfer calculation to determine the dust temperature at each location of the disk. The radiative transfer calculation is performed with the 3D Monte Carlo radiative transfer code, \textsc{radmc-3d} \footnote{The code and more information are available on \url{http://www.ita.uni-heidelberg.de/~dullemond/software/radmc-3d/}}. The number of grids is $N_r=256$ for $R_{\rm in}<R<R_{\rm out}$, $N_\phi=256$ for $0<\phi<2\pi$, and $N_\theta=128$ for $\pi/3<\theta<2\pi/3$ (Figure \ref{fig:grid}). The number of photons is set as $10^9$.

In the radiative transfer calculation, dust grains are assumed to be spherical, and this allows us to use an exact solution for the optical properties from the Mie theory \citep[e.g.,][]{Bohren:1983aa}. In addition, we average over the optical properties with respect to the power-law size distribution,
\begin{equation}
n(a)da \propto a^{p}da\ (a_{\rm min}<a<a_{\rm max})
\end{equation}
where $n(a)da$ is the number of spherical dust grains having sizes of $a$ to $a+da$.
We assume the grains are well mixed in the gas density distribution.
We adopt the following parameters for the grain size distribution: $a_{\rm min}=0.005\ \mu$m, $a_{\rm max}=100\ \mu$m, $p=-3.5$.
The dielectric function of the grains is assumed to be a mixture of the silicate and the H$_2$O ice \citep{Miyake:1993aa}.

Based on the temperature structure determined by \textsc{radmc-3d}, the energy density of radiation field at each location is calculated by
\begin{equation}
u_{{\rm rad}, \nu}(r,\theta)= \frac{1}{c}\left\{\int_0^{2\pi} I_\nu(\alpha)d\alpha + B_{\nu}(T_*)\Delta\Omega_*e^{-\tau_{\nu, *}}\right\}
\end{equation}
where $\Delta\Omega_*=\pi(R_*/r)^2$ denotes the solid angle of the star, $r$ is the distance from the star, and $\tau_{\nu, *}$ is the optical depth along the line of sight of the central star. The intensity at the $i$th grid, $I_\nu^i(\alpha)$, is calculated by solving
\begin{equation}
I_\nu^{i}(\alpha)=B_\nu(T_d^{i})(1-e^{-\delta\tau_i})+\sum_{j\neq i} B_\nu(T_d^{j})(1-e^{-\Delta\tau_j})e^{-\tau_{ij}} \label{eq:intensity}
\end{equation}
where $T_d^j$ is the dust temperature on the $j$th grid, $\delta\tau^i$ is the optical depth from the $i$th grid center to the $i$th grid wall, $\tau_{ij}$ is the optical depth of the $i$th and the $j$th grid wall, and $\Delta\tau_j$ is the optical depth of the $j$th grid. The summation is performed along the line of sight. 
We neglect the scattered starlight because at the surface layer RAT is dominated by the direct emission from the star. \footnote{If dust grains at the surface layer have a large albedo, scattered starlight may dominate the disk mid-infrared polarization, and polarized thermal emission arising from aligned grains becomes subdominant. In this case, we expect an azimuthal polarized vector for face-on disks. Note that even in this case, we expect circular polarization from the disk as discussed in Section \ref{sec:V}}
The anisotropy parameter $\gamma_\nu$ is calculated approximately by
\begin{equation}
\gamma_\nu\approx\frac{I_\nu(\alpha=180^{\circ})-I_\nu(\alpha=0^{\circ})}{I_\nu(\alpha=180^{\circ})+I_\nu(\alpha=0^{\circ})}
\end{equation}

\subsection{A magnetic field strength and topology} \label{sec:bfield}
The magnetic field strength is one of the most uncertain quantities in protoplanetary disks.
Global magnetohydrodynamics (MHD) simulation of disks suggests that the magnetic field is dominated by the toroidal component because of the disk shear flow \citep[e.g.,][]{Flock:2015aa}. Hence, we only consider the toroidal magnetic field.
\citet{Okuzumi:2014aa} studied the radial transport of the magnetic field and estimated an upper limit on the vertical field strength. They found that the maximum vertical field strength is
\begin{equation}
B_{\rm z,\ max}(R) = 1\ {\rm mG}\left(\frac{R}{10\ {\rm au}}\right)^{-2}.
\end{equation}
The toroidal magnetic field can be amplified up to 10 times larger than the initial vertical magnetic field strength. 
However, $B_{\rm z,\ max}(R)$ indicates the maximum value of the initial vertical $B$-field strength, and the actual strength may be smaller; hence, we assume $B_{\rm z,\ max}(R)$ as the toroidal $B$-field strength.

\section{Grain alignment in the disk} \label{sec:result1}
\subsection{Anisotropic radiation field in the disk} \label{sec:radfield}
\begin{figure*}[t]
\begin{center}
\includegraphics[height=5cm,keepaspectratio]{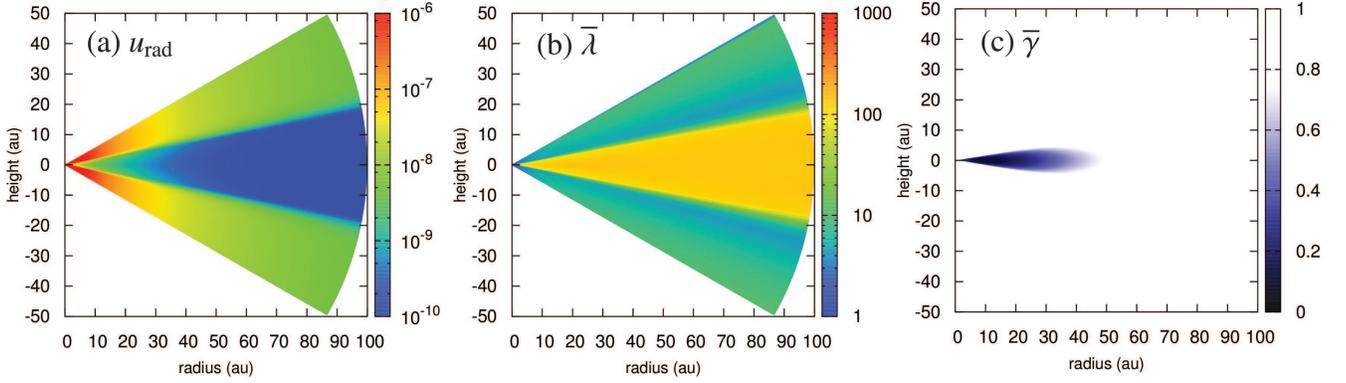}
\caption{The radiation field in the disk. (a) Energy density of the radiation $u_{\rm rad}$ [erg cm$^{-3}$] (b) the spectrum averaged wavelength $\overline{\lambda}$ [$\mu$m], and (c) the spectrum averaged anisotropy parameter $\overline{\gamma}$ are shown.}
\label{fig:radfield}
\end{center}
\end{figure*}

In Figure \ref{fig:radfield}, we show the energy density of the radiation field, the spectrum weighted wavelength, and the anisotropy parameter in the disk.
Above the disk photosphere, the energy density of radiation is dominated by the stellar radiation. Hence, at the surface layer, $u_{\rm rad}\approx \mathcal{A}T^4\Delta\Omega_*$, where $\mathcal{A}=7.57\times10^{-15}$ erg cm$^{-3}$ K$^{-4}$ is the radiation constant, and then
\begin{equation}
u_{\rm rad}\simeq 5.2\times10^{-6}\ {\rm erg\ cm}^{-3} \left(\frac{R}{10\ {\rm au}}\right)^{-2} \label{eq:uradsur}
\end{equation}
for $T_*=4000$ K and $R_*=2R_{\odot}$.
The anisotropy parameter of the radiation field is almost unity, and $\vecbf{k}$ is pointing radially outward.

At the disk midplane, the energy density of the radiation is $2-3$ orders of magnitude lower than that at the surface layer because the disk is optically thick at visible/NIR wavelengths. The energy density of the radiation field at the midplane is dominated by the thermal emission from cold grains at the midplane ($\bar{\lambda}\simeq 140\ \mu$m). Even in the midplane, the radiation energy density is still very large compared to the interstellar radiation field, $u_{\rm ISRF}=8.64\times 10^{-13}$ erg cm$^{-3}$ \citep{Mathis:1983aa}. As shown in Figure \ref{fig:radfield}(c), the anisotropy parameter becomes around $0.1$ at the midplane at $R\lesssim 30$ AU because the disk becomes optically thick for the radiation with $\overline{\lambda}\approx140 \mu$m. Therefore, in this region, only 10 \% of the energy density can contribute the RAT because only the anisotropic radiation field is important for RAT alignment. At the outer region, the disk becomes optically thin, and the radiation anisotropy is almost 100 \% and it coincides with the radial direction. It is worth noting that when we consider a dust ring structure with an inner hole around the central star, the radiation anisotropy can be in the azimuthal direction \citep{kataoka2015}. Dust grains may become aligned with the azimuthal radiative flux if the ring is not uniform along the azimuthal direction. In this paper, we assume a smooth disk density distribution from the region near the central star to the outer disk; thereby, the anisotropic radiation directs radially outward.
 
\subsection{RAT alignment timescale in the disk}
\begin{figure*}[t]
\begin{center}
\includegraphics[height=11.0cm,keepaspectratio]{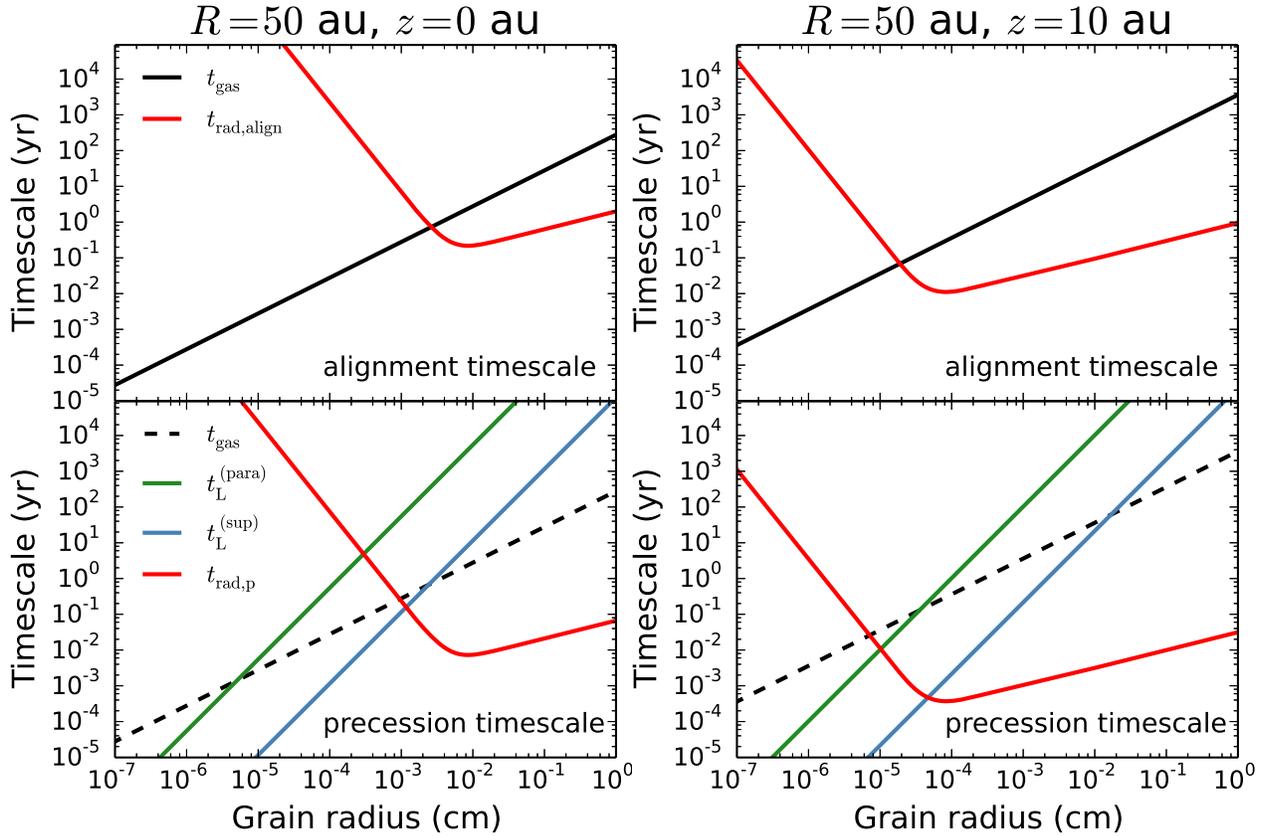}
\caption{Timescales relevant to RAT alignment. Top and bottom panels show the alignment timescales and the precession timescales, respectively.
Left and right panels correspond to the two different locations in the disk, $(R,z)=(50\ \rm{au}, 0\ \rm{au})$, and $(R,z)=(50\ \rm{au}, 10\ \rm{au})$, respectively.
For the top panels, black and red lines represent the timescale of the gaseous damping ($\tgas$) and the RAT alignment timescale ($\tradalign$). 
For the bottom panels, the red solid line indicates the radiative precession timescale ($\tradp$). Green and blue lines show the Larmor precession timescale ($\tl$) for paramagnetic inclusions ($f_p=10$\%) and superparamagnetic inclusions ($f_p=10$\%, $\phi_{\rm sp}=3$\%, $N_{\rm cl}=2\times10^3$), respectively. The dashed line in the bottom panels represents the gaseous damping timescale.}
\label{fig:timescales}
\end{center}
\end{figure*}

Based on the radiation field calculated in Section \ref{sec:radfield}, we calculate the characteristic timescales relevant to grain alignment in the disk.
Figure \ref{fig:timescales} shows the timescales relevant to RAT alignment at two different locations (midplane and surface at $R=50$ au) in the disk.

The alignment timescale shown in the top panels in Figure \ref{fig:timescales} indicates the minimum size of dust grains to be aligned.
In general, smaller grains experience stronger gaseous damping and also the RAT becomes less efficient; therefore, only large grains can be aligned.
A critical grain size is given by the balance between the RAT alignment timescale ($\tradalign$) and the gaseous damping timescale ($\tgas$).
At the midplane of $R=50$ au, the RAT is determined by the radiation field emitted from ambient thermal grains with $\overline{\lambda}\approx140 \mu$m, and then for smaller grains, $a \leq 140\ \mu{\rm m}/1.8 \approx 80 \mu$m, the RAT becomes less efficient (see Equation \ref{eq:QRATs}).
Figure \ref{fig:timescales} shows that at the midplane of $R=50$ au, the critical grain size is $\approx 27\ \mu$m.
As a result, external alignment for a mm-sized grain is possible even at the dense midplane, although these grains may not show internal alignment (see Figure \ref{fig:internal}). At the surface layer, the gas density is much lower than the that at the midplane, and the radiation field has a shorter wavelength than the midplane; hence, external alignment of smaller grains is possible. Figure \ref{fig:timescales} shows the critical grain size at the disk surface is $\approx 0.2\mu$m. 

The bottom panels in Figure \ref{fig:timescales} show the precession timescales of dust grains, and these define the alignment axis.
As we mentioned in Section \ref{sec:alignment}, the alignment axis is determined by the dominant precession motion of the grains.
Figure \ref{fig:timescales} shows that, at the midplane, a dust grain which is larger than the critical size precesses around the radiative flux; therefore, these grains become aligned with radiation direction. Even if we consider superparamagnetic inclusions, Larmor precession is still slow compared to radiative precession. Therefore, these grains become aligned with respect to the radiation direction, not to the magnetic field. Note that the direction of radiation at the midplane of $R=50$ au is radially outward; hence, the aligned grains are expected to produce azimuthally polarized emission. 
In addition to the fast radiative precession, it is also found that Larmor precession is often suppressed by the gaseous damping. The presence of superparamagnetic inclusions enhances the magnetic susceptibility significantly; nevertheless, only grains smaller than $\approx 10\ \mu$m can overcome the gaseous damping. This result also illustrates the difficulty of grain alignment with respect to the magnetic field in the disk midplane. 
A critical grain size for which the Larmor precession timescale equals to the gaseous damping can be obtained from Equations (\ref{eq:tgas}) and (\ref{eq:tl}):
\begin{equation}
a_{\rm crit}\simeq0.1\ \mu{\rm m}\ \hat{\rhog}^{-1}\hat{T_g}^{-1/2}\hat{\chi}\left(\frac{B}{1\ {\rm mG}}\right). \label{eq:acrit}
\end{equation}
If the grain size is larger than this critical size, Larmor precession is suppressed by the gaseous damping. 
In summary, (sub-)mm-sized dust grains at the midplane are difficult to align with the magnetic field due to the efficient gaseous damping as well as fast radiative precession; however, these grains may align with the radiation direction.

At the surface layer, a grain size larger than 0.2 $\mu$m can be aligned.
Figure \ref{fig:timescales} shows that sub-micron-sized grains can be aligned with the magnetic field, while larger grains become aligned with the radiation direction. These grains may contribute to the mid-infrared polarization of the disk because the disk is optically thick at mid-infrared wavelength, and then the surface grains are responsible for the polarized flux. Although sub-micron-sized grains without superparamagnetic inclusions align with the radiation direction at surface layer, those with superparamagnetic inclusions can become aligned with the magnetic field. As a result, in mid-infrared polarimetry, we expect magnetic field alignment for grains having superparamagnetic inclusions. 

Finally, it is worth mentioning how carbonaceous grains affect on grain alignment.
As shown in \citet[][]{LH07}, the alignment induced with respect to the radiation direction is mostly independent of the grain composition, while the alignment with respect to the magnetic field depends on the Larmor precession frequency, which is significantly reduced for 
carbonaceous grains. As a result, the introduction of the carbonaceous grains results in increasing the radius over which the grains are aligned with respect to the radiation direction.

\section{Degree of polarization of the disk} \label{sec:result2}
We calculate the polarization flux arising from aligned grains, and estimate the degree of polarization of the disk.
To focus on how grain alignment affects on the degree of polarization, we neglect the polarization due to the scattering.
In addition, we assume a face-on disk for the sake of simplicity; the effect of disk inclination is discussed in Section \ref{sec:inclination}.
In Section \ref{sec:ellipsoid}, we describe an analytical model of the degree of linear polarization arising from a single ellipsoid.
In Section \ref{sec:diskpolarization}, we present a model of the disk polarization with/without dust settling.
The results are presented in Section \ref{sec:nomag}, \ref{sec:withmag}, and \ref{sec:sizedep}.

\subsection{Linear degree of polarization of an ellipsoid} \label{sec:ellipsoid}
\begin{figure}[b]
\begin{center}
\includegraphics[height=7cm,keepaspectratio]{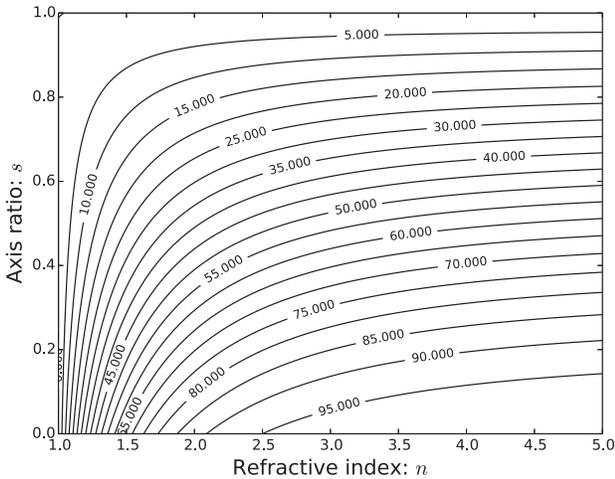}
\caption{Degree of polarization arising from an oblate spheroid with an axis ratio of $s$ and a refractive index of $n$ in the Rayleigh limit, calculated using Equations (\ref{eq:optaxis}) and (\ref{eq:polari}).}
\label{fig:p}
\end{center}
\end{figure}

\citet{CL07} calculated the absorption cross section along the minor axis and the major axes using the DDSCAT code \citep{draine94}; the degree of polarization is obtained by
\begin{equation}
p=\frac{C_{{\rm abs},\perp}-C_{{\rm abs},||}}{C_{{\rm abs},\perp}+C_{{\rm abs},||}} \label{eq:polari}
\end{equation}
where $C_{{\rm abs},\perp}$ and $C_{{\rm abs},||}$ are the absorption cross section for $\vecbf{E}\perp \hat{\vecbf{a_1}}$ and $\vecbf{E} || \hat{\vecbf{a_1}}$, respectively, where $\vecbf{E}$ represents the electric field vector of the light. Obviously, for a spherical grain, $C_{{\rm abs},\perp}=C_{{\rm abs},||}$; therefore, unpolarized thermal emission is radiated. \citet{CL07} showed that the degree of polarization becomes almost zero when a grain radius is larger than $\lambda/2\pi$, e.g., the geometrical optics limit.
Hence, we assume
\begin{equation}
 p(a_{\rm eff}, \lambda)\approx \left\{ \begin{array}{ll}
 \frac{C_{{\rm abs},\perp}-C_{{\rm abs},||}}{C_{{\rm abs},\perp}+C_{{\rm abs},||}}, 
 & 2\pi{a_{\rm eff}}<\lambda \\[3mm]
 0,
 & {\rm otherwise}.
\end{array} \right.
\label{eq:pmodel}
\end{equation}
For $2\pi{a_{\rm eff}}<\lambda$, we use an electrostatic analysis of the ellipsoid (Rayleigh approximation) to find the value of $C_{{\rm abs},\perp}$ and $C_{{\rm abs},||}$ \citep[e.g.,][]{Bohren:1983aa}.
Suppose $\alpha_j$ is the polarizability of the ellipsoid with respect to the axis $j$, where $j$ runs from 1 to 3; then it becomes
\begin{equation}
\alpha_j=4\pi a_1a_2^2\frac{m^2-1}{3+3L_j(m^2-1)}, \label{eq:alpha}
\end{equation}
where $L_j$ is a geometrical parameter and $m$ is the complex refractive index. 
In the case of an oblate ellipsoid, the geometrical parameter satisfies $L_1=L_2<L_3$ and $L_1+L_2+L_3=1$.
In this case, $L_1$ becomes
\begin{eqnarray}
L_1&=&\frac{g(e)}{2e^2}\left[\frac{\pi}{2}-\arctan(g(e))\right]-\frac{g(e)^2}{2}\\
g(e)&=&\left(\frac{1-e^2}{e^2}\right)^{1/2}
\end{eqnarray}
where $e^2=1-s^2$. 
The absorption cross section is given by  the optical theorem, $C_{\rm abs}=k{\rm Im}(\alpha_j)$, where $k$ is a wavenumber. 
Denoting the complex refractive index by $m=n+i{\rm k}$\footnote{The symbol $k$ is used to denote the wavenumber, while the imaginary part of the complex refractive index is denoted by ${\rm k}$ according to \citet{Bohren:1983aa}.}, and assuming a less absorbing grain (${\rm k} \ll 1$), we obtain
\begin{eqnarray}
C_{\rm abs,j}\simeq\frac{2\pi}{\lambda}V\frac{2n{\rm k}}{[L_j(n^2-1)+1]^2}
\end{eqnarray}
where $V$ is the volume of an ellipsoid. Finally, we obtain
\begin{equation}
\frac{C_{{\rm abs},\perp}}{C_{{\rm abs},||}}=\frac{[L_3(n^2-1)+1]^2}{[L_1(n^2-1)+1]^2} \label{eq:optaxis}
\end{equation}
where $L_3=1-2L_1$. Since the geometrical parameter $L$ depends only on the axis the ratio of the grains, Equation (\ref{eq:optaxis}) is irrelevant to the grain size $\aeff$.
In Figure \ref{fig:p}, we plot Equation (\ref{eq:polari}) for a Rayleigh approximation. 
When we assume astronomical silicate \citep{draine84, laor93} as the dielectric function, and adopt $\lambda=850 \mu$m, Equation (\ref{eq:polari}) gives $C_{{\rm abs},\perp}/C_{{\rm abs},||}=1.6$ for $s=0.77$, and hence the degree of polarization is 22\%.
For $s=0.67$, the ratio is $C_{{\rm abs},\perp}/C_{{\rm abs},||}=2.1$, and hence, the degree of polarization is 35\%.
These results are compatible with the DDA calculations of \citet{CL07}.

\subsection{Polarized emission from the disk} \label{sec:diskpolarization}
Combining with the temperature structure obtained by using \textsc{radmc-3d} and the degree of polarization of a single oblate grain (Equation \ref{eq:polari}), we  can calculate the polarized flux from the disk.

The disk model we presented in Section \ref{sec:diskmodel} assumed a well-mixed population of dust grains with the gas at everywhere in the disk.
In other words, the grain size distribution is the same in every grid cell in the disk.
However, this assumption may lead to an underestimation of the degree of polarization at mid-infrared wavelengths. This is because the presence of dust grains larger than the observing wavelength at the surface layer reduces the polarization degree (Equation \ref{eq:pmodel}), while such large grains are expected to be depleted from the surface layer due to dust settling. 
Hence, we consider dust settling in the calculation of the polarized emission in order to study its effect on the results.
To reduce the computational cost, we adopt the following assumptions in the ray-tracing procedure. Firstly, the temperature of each grid cell in the disk is assumed to be the same as in Section \ref{sec:diskmodel}. Secondly, we calculate the emissivity on each grid cell taking dust settling into account (see Section \ref{sec:distribution} for more details). Thirdly, for a given emissivity of each grid cell, the ray-tracing calculation is performed.

\subsubsection{Spatial and size distribution model of grains} \label{sec:distribution}
We assume the grain surface number density in a size range [$a, a+da$], $\mathcal{N}(R,a)$, obeys a power-law with power-law index $-3.5$; then
\begin{equation}
\mathcal{N}(R,a)da=\mathcal{N}_0(R)a^{-3.5}da\ (a_{\rm min}<a<a_{\rm max})
\end{equation}
Since 
\begin{equation}
\Sigma_d(R)=\int_{a_{\rm min}}^{a_{\rm max}} \frac{4}{3}\pi a^3\rho_s\mathcal{N}(R,a)da,
\end{equation}
we obtain the radial distribution of the grain surface number density,
\begin{equation}
\mathcal{N}_0(R)=\frac{3}{8\pi}\frac{\Sigma_d}{\rho_s}\left[a_{\rm max}^{1/2}-a_{\rm min}^{1/2}\right]^{-1}
\end{equation}
We simply assume a Gaussian density profile in the vertical direction; then the dust density in a size range [$a, a+da$] can be written by
\begin{equation}
\rho_d(R,z,a)da = \rho_d(R,0,a)\exp \left[-\frac{z^2}{2h_d^2}\right]da \label{eq:rhod_rza}
\end{equation}
where $h_d$ is the dust scale height. Assuming that this is determined by the balance between gravitational settling and turbulent diffusion \citep{Youdin:2007aa},  we have
\begin{equation}
h_d=h_g\left(1+\frac{\Omega t_s}{\alpha}\frac{1+2\Omega t_s}{1+\Omega t_s}\right)^{-1/2} \label{eq:hd}
\end{equation}
where $\Omega$ is the Kepler angular velocity, $\alpha$ is the turbulent strength, $h_g$ is the gas pressure scale height, and $t_s$ is the stopping time given by
\begin{equation}
 t_s = \left\{ \begin{array}{ll}
 t_s^{\rm(Ep)} \equiv \frac{3m}{4\rho_g v_{\rm th} A}, 
 & a< \frac{9}{4}\lambda_{\rm mfp}, \\[3mm]
 t_s^{\rm(St)} \equiv \frac{4a}{9\lambda_{\rm mfp}} t_s^{\rm(Ep)},
 & a> \frac{9}{4}\lambda_{\rm mfp}, 
\end{array} \right.
\label{eq:ts}
\end{equation}
where (Ep) and (St) indicate the Epstein and Stokes drag laws, respectively, $m$ is the mass of the grain, $A$ is the cross section of the grain, and $\lambda_{\rm mfp}=\mu m_{\rm H}/\sigma_{\rm mol}\rho_g$ is the mean free path of a gas molecule, where we adopt $\sigma_{\rm mol}=2\times10^{-15}$ cm$^{2}$ as the collisional cross section of a molecule.
By integrating Equation (\ref{eq:rhod_rza}) with respect to $z$, we can obtain
\begin{equation}
\rho_d(R,0,a)=\frac{1}{2}\frac{\Sigma_d}{\sqrt{2\pi}h_d}\left[a_{\rm max}^{1/2}-a_{\rm min}^{1/2}\right]^{-1}a^{-\frac{1}{2}}
\end{equation}
As a result, the number density of dust grains in the size range [$a, a+da$] is
\begin{eqnarray}
n(R,z,a)da&=&\frac{3\rho_d(R,z,a)}{4\pi a^3\rho_s}da\\
&=&\frac{3}{8\pi}\frac{\Sigma_d/\rho_s}{\sqrt{2\pi}h_d}\left[a_{\rm max}^{1/2}-a_{\rm min}^{1/2}\right]^{-1}a^{-3.5}\exp \left[-\frac{z^2}{2h_d^2}\right]da \nonumber \\
\end{eqnarray}

\subsubsection{Polarization degree of emission of grains with size distribution} \label{sec:diskpol}
To calculate the emissivity, we define a locally averaged grain opacity, $\langle \kappa_{\rm abs,sca} \rangle (R,z)$ by
\begin{equation}
\langle \kappa_{\rm abs} \rangle (R,z)=\frac{\int_{a_{\rm min}}^{a_{\rm max}}n(R,z,a)a^3\kappa_{\rm abs}(a)da}{\int_{a_{\rm min}}^{a_{\rm max}}n(R,z,a)a^3da} \label{eq:opcave}
\end{equation}
At each location in the disk, the intrinsic degree of polarization is
\begin{equation}
\langle p_0(R,z,\lambda) \rangle= 
\frac{\int_{a_{\rm min}}^{a_{\rm max}} (-1)^nP(a,\lambda)\mathcal{R}(a)\kappa_{\rm abs}(a,\lambda)a^3n(R,z,a)da}{\int_{a_{\rm min}}^{a_{\rm max}} \kappa_{\rm abs}(a,\lambda)a^3n(R,z,a)da}. 
\label{eq:plocal}
\end{equation}
where we insert $(-1)^n$ so that the direction of polarization is considered. We assume that the polarization degree becomes negative ($n=1$) for a radial $\vecbf{E}$-vector ($\tl/\tradp<1$) and positive ($n=2$) for an azimuthal $\vecbf{E}$-vector ($\tl/\tradp>1$). 
$n$ and $\mathcal{R}(a)$ are calculated with the radiation field given in Figure \ref{fig:radfield}. The absorption opacity in Equation (\ref{eq:plocal}) is calculated by using the Mie theory, regarding an oblate as a sphere with $\aeff$.

Plugging Equation (\ref{eq:pmodel}) into Equation (\ref{eq:plocal}), we obtain
\begin{eqnarray}
\langle p_0(R,z,\lambda) \rangle&=& 
\frac{C_{{\rm abs},\perp}-C_{{\rm abs},||}}{C_{{\rm abs},\perp}+C_{{\rm abs},||}}\times \nonumber \\
&&\frac{\int_{a_{\rm min}}^{a_{\rm large}} (-1)^n\mathcal{R}(a)\kappa_{\rm abs}(a,\lambda)a^3n(R,z,a)da}{\int_{a_{\rm min}}^{a_{\rm max}} \kappa_{\rm abs}(a,\lambda)a^3n(R,z,a)da} \nonumber \\
\label{eq:pave}
\end{eqnarray}
where $a_{\rm large}\equiv\min(a_{\rm max}, \lambda/2\pi)$ represents the maximum size of dust grains being aligned.
In Equation (\ref{eq:pave}), we assume for simplicity that $a_2/a_1$ is constant through the grain size distribution and all grains have the same temperature.

Using Equations (\ref{eq:opcave}) and (\ref{eq:pave}), we calculate the polarized intensity and the total intensity via the ray-tracing method (Figure \ref{fig:grid}). The polarization degree is then obtained as the polarized intensity divided by the intensity at each radius of the disk.

\subsection{Alignment of grains without magnetic inclusions} \label{sec:nomag}
\begin{figure*}[htbp]
\begin{center}
\includegraphics[height=8.0cm,keepaspectratio]{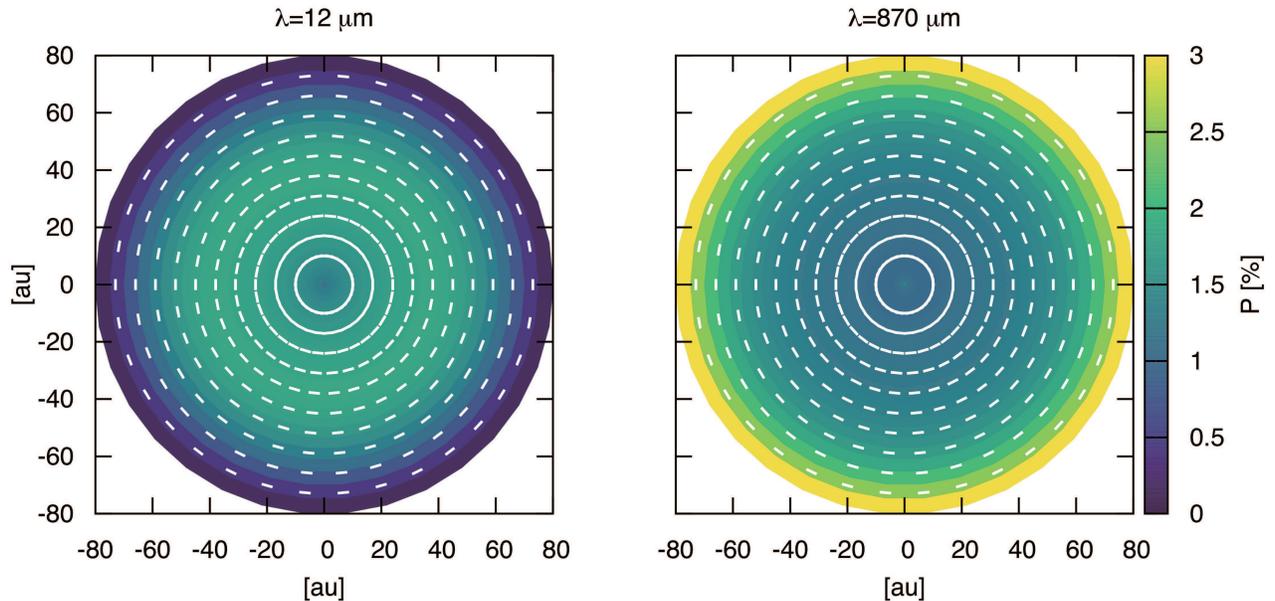}
\caption{Polarization degree is shown with the color scale and the direction of $\vecbf{E}$-vector is plotted as the white bar. Left and right panels represent mid-infrared wavelength ($\lambda=12\ \mu$m) and millimeter wavelength ($\lambda=870\ \mu$m.), respectively. The dust grains are assumed to be magnetically poor ($f_p=0.01$ and $\phi_{\rm sp}=0$). The maximum grain size is $a_{\rm max}=1000 \mu$m. Turbulent strength and the $f_{{\rm high-}J}$-parameter are assumed to be $\alpha=10^{-3}$ and $f_{{\rm high-}J}=0.5$, respectively.}
\label{fig:ID15}
\end{center}
\end{figure*}

In Figure \ref{fig:ID15}, we plot a map of the degree of polarization of the disk with a polarization vector at mid-infrared wavelength ($\lambda=12\ \mu$m) and millimeter wavelength ($\lambda=870\ \mu$m). The dust grains are assumed to be magnetically poor ($f_p=0.01$ and $\phi_{\rm sp}=0$) and the disk magnetic field model is given in Section \ref{sec:bfield}.
Figure \ref{fig:ID15} shows that for both mid-infrared and submillimeter wavelengths, the polarization vector shows a centrosymmetric pattern. This is because the dust grains align with the radiation direction (see also Figure \ref{fig:timescales}). Therefore, even in the presence of a toroidal magnetic field, the polarization vector can be in the azimuthal direction for both mid-infrared wavelength and millimeter wavelengths. 

\begin{figure}[htbp]
\begin{center}
\includegraphics[height=6.0cm,keepaspectratio]{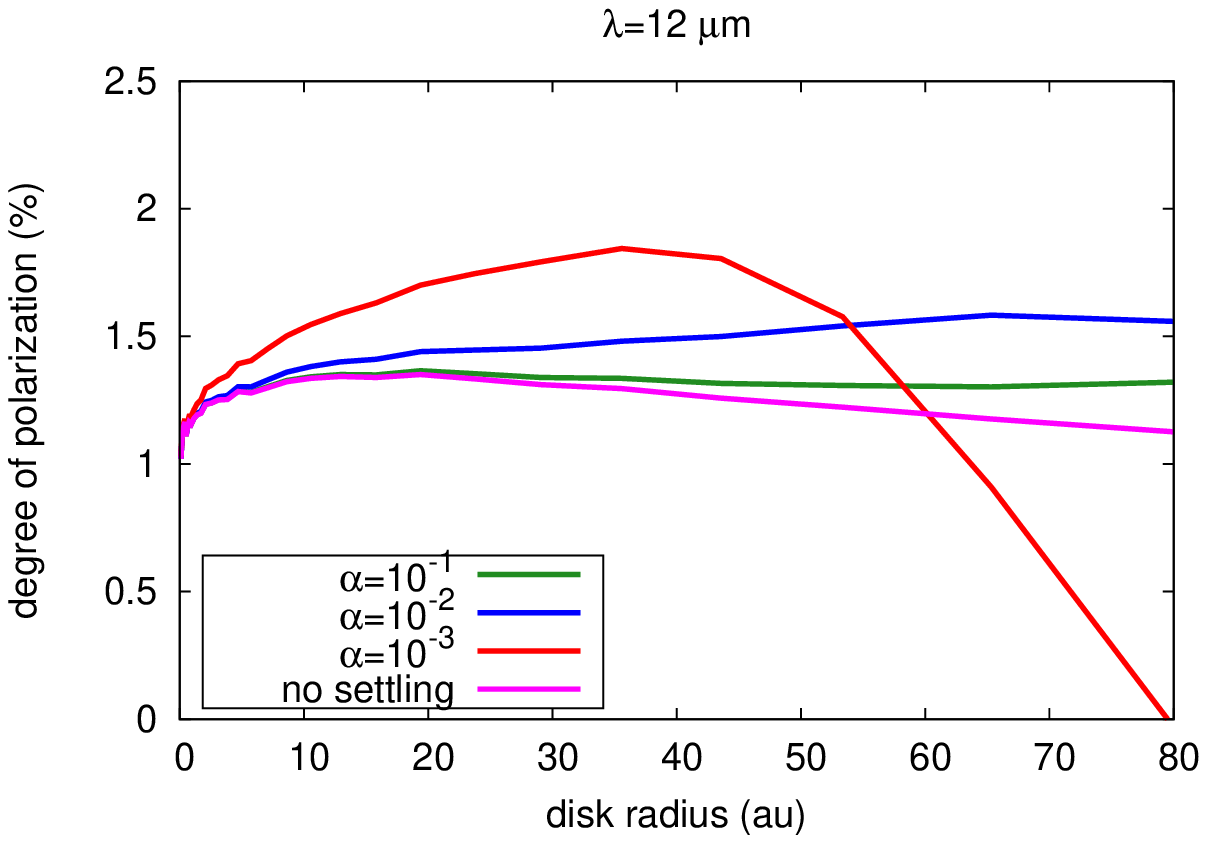}
\includegraphics[height=6.0cm,keepaspectratio]{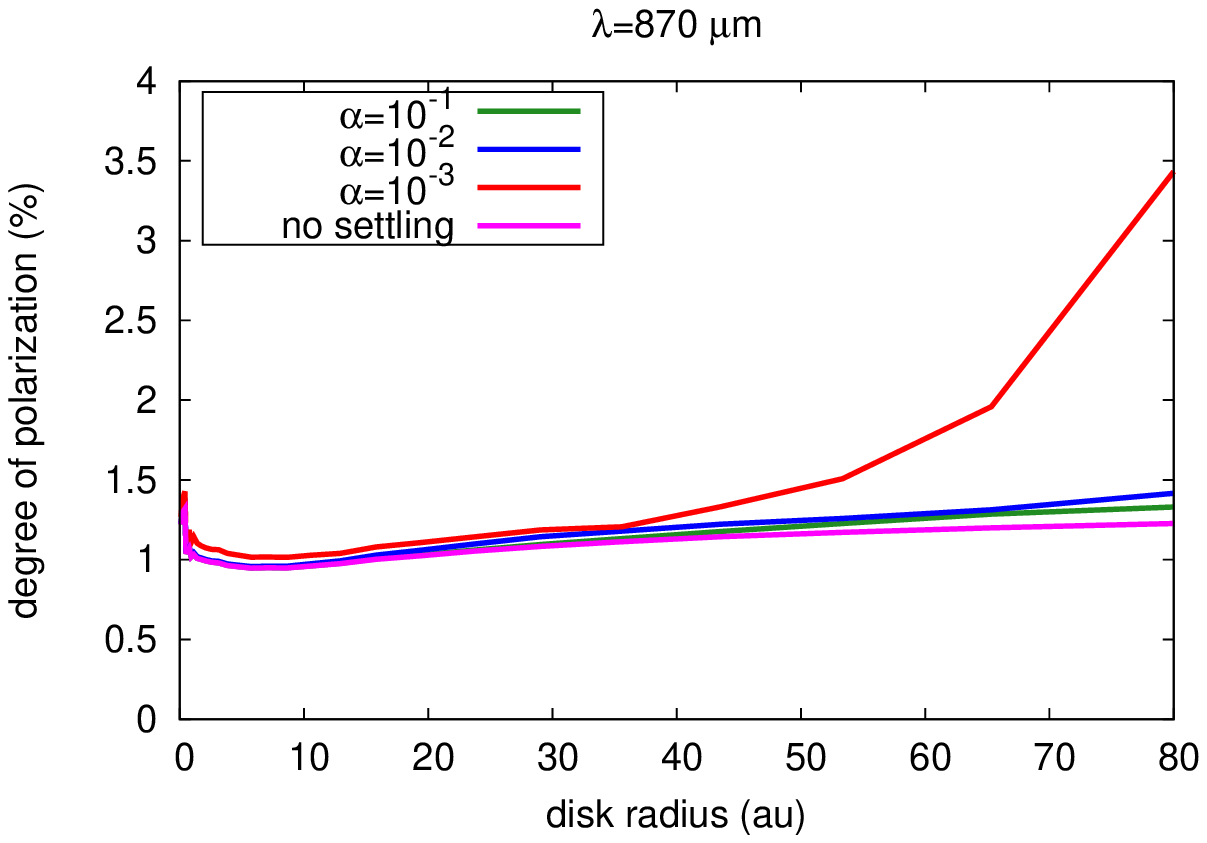}
\caption{Degree of polarization against disk radius. The turbulent strength is varied, but the other parameters are the same as Figure \ref{fig:ID15}. Top and bottom panels indicate mid-infrared wavelength ($\lambda=12\ \mu$m) and millimeter wavelength ($\lambda=870\ \mu$m), respectively. Red, blue, and green lines represent the results for different $\alpha$-values, $\alpha=10^{-3},10^{-2}$, and $10^{-1}$, respectively, and the magenta line represents the no-settling model.}
\label{fig:P_radial}
\end{center}
\end{figure}

The radial dependence of the degree of polarization differs depending on dust settling, or the turbulent strength, as shown in Figure \ref{fig:P_radial}. 
For the case of no dust settling, the radial dependence is not strong for both of mid-infrared and millimeter wavelengths. 

At mid-infrared wavelength with $\alpha=10^{-1}$ and $10^{-2}$, the degree of polarization increases with increasing the disk radius, and at the inner region, it coincides with that of the no-settling case. This is because at the inner region, the gas density is sufficiently high that the largest grains can be well coupled to the gas. On the other hand, at the outer region, the largest grains settle down to the midplane because of the low gas density. Since grains larger than the observing wavelength reduce the degree of polarization, their depletion from the surface layer can enhance the mid-infrared polarization degree.
If we consider a disk with weak turbulence, $\alpha=10^{-3}$, the degree of polarization of the outer disk decreases with increasing the disk radius. Under very weak turbulence, micron-sized grains also settle down, and only sub-micron-sized grains can be stirred up to the surface layer. However, sub-micron-sized metal poor grains do not show external alignment, as shown in Figure \ref{fig:timescales}. As a result, dust settling leads to a reduction of the polarization degree. 
Note that once micron-size dust grains settle down to the midplane, they hardly contribute to mid-infrared polarization because the gas density at the midplane is high, so that the gaseous damping timescale is short, and the wavelength of the radiation field is long, and so RAT is inefficient for such grains. Therefore, these grains are not likely to be aligned.

\citet{Li:2016aa} concluded that the gradual increase of polarization degree is due to the scattering. Our result shows that in the presence of dust settling with a moderate strength of turbulence may also produce a monotonically increasing radial dependence of the degree of polarization.

At the millimeter wavelength region, the radial dependence does not change significantly except for the case of $\alpha=10^{-3}$. 
For a low-turbulent disk ($\alpha=10^{-3}$), efficient dust settling results in the formation of a thin layer of large particles at the midplane, i.e., $a\geq \lambda/2\pi\simeq 140\ \mu$m; hence, at the region where just above the midplane, but well below the photosphere, large grains can be depleted. As a result, grain sizes smaller than $\lambda/2\pi\simeq 140\ \mu$m at just above the thin midplane layer can produce the polarized flux. However, the formation of the thin dust layer may be halted by the Kelvin-Helmholtz instability; then the strong radial dependence for $\alpha=10^{-3}$ may not appear in a realistic disk.

\subsection{Alignment of grains with magnetic inclusions} \label{sec:withmag}
\begin{figure*}[htbp]
\begin{center}
\includegraphics[height=8.0cm,keepaspectratio]{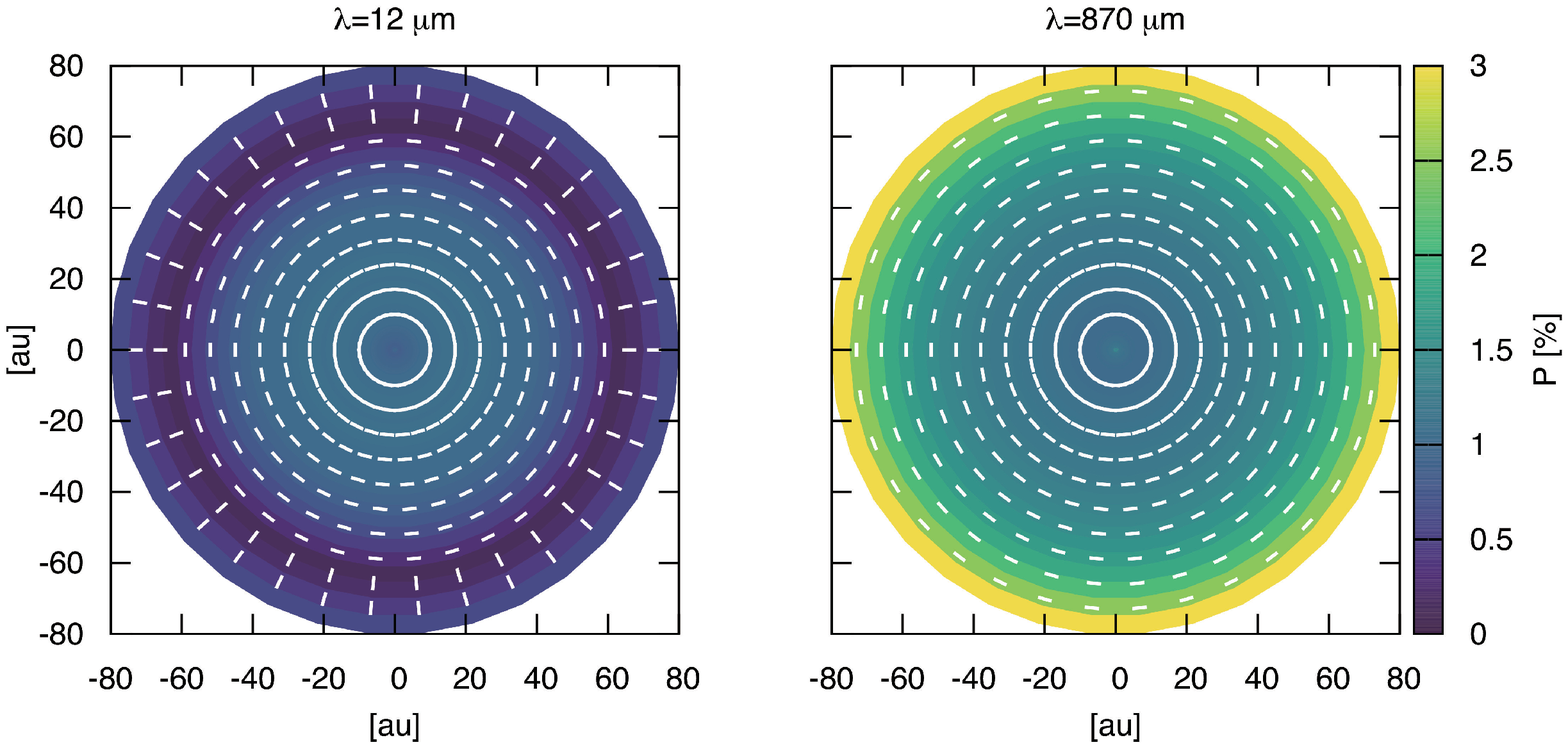}
\includegraphics[height=8.0cm,keepaspectratio]{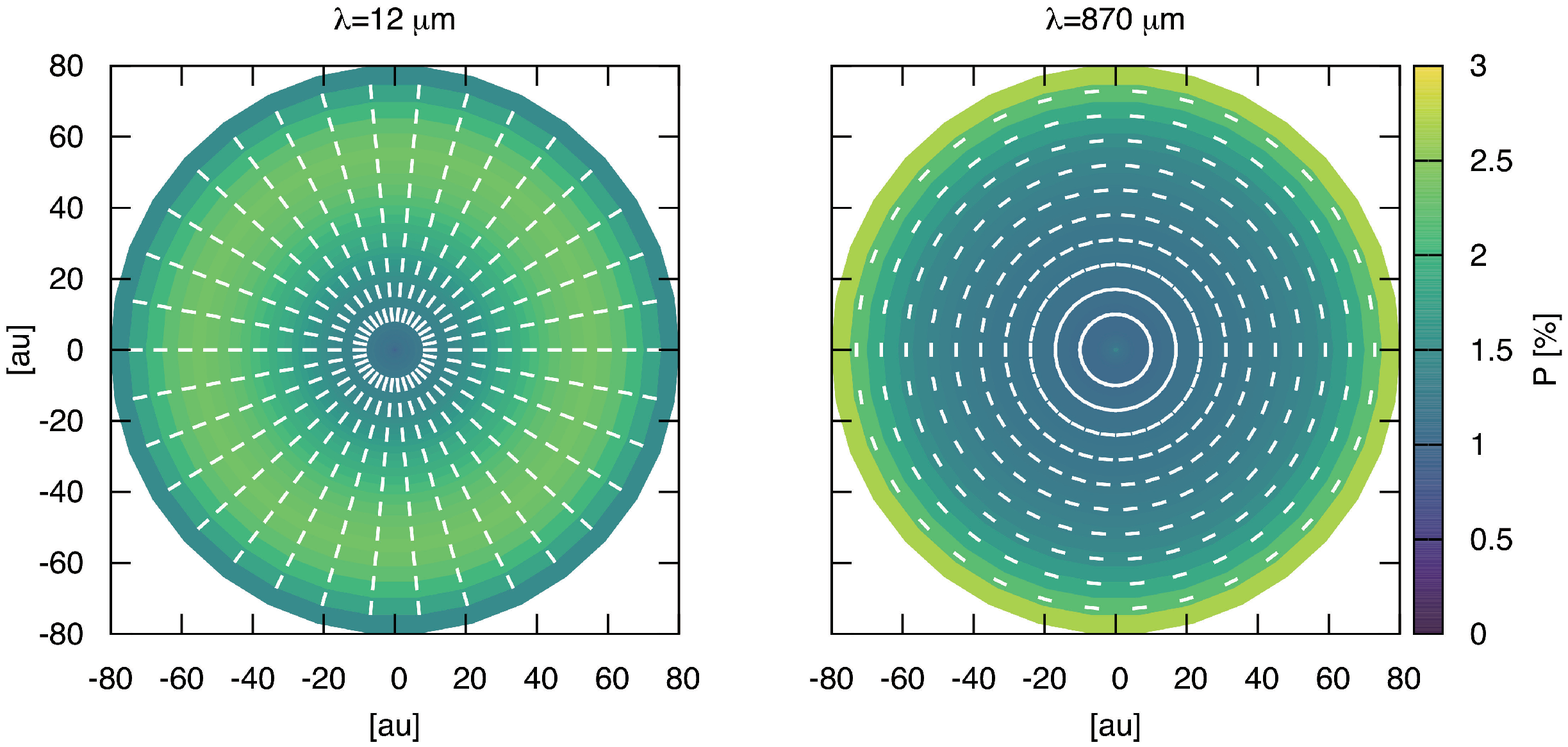}
\caption{Same as Figure \ref{fig:ID15}, but for different magnetic properties of dust. Top and bottom panels correspond to (i) moderate paramagnetic inclusion model ($f_p=0.1$, $\phi_{\rm sp}=0$), and (ii) superparamagnetic inclusion model ($f_p=0.1$, $\phi_{\rm sp}=0.03$, $N_{\rm cl}=2000$), respectively. $a_{\rm max}=10^{3}\ \mu$m, $f_{\rm high-J}=0.5$, and $\alpha=10^{-3}$ are assumed.}
\label{fig:imagemap}
\end{center}
\end{figure*}

We investigate how does the presence of magnetic inclusions affects on grain alignment. 
Figure \ref{fig:imagemap} is the same as Figure \ref{fig:ID15}, but for a different number of magnetic inclusions. In the presence of a modest number of magnetic inclusions and no superparamagnetic inclusions ($f_p=0.1$, $\phi_{\rm sp}=0$), the figure shows grain alignment with radiative flux at the inner disk, and with the magnetic field at the outer disk at mid-infrared wavelength. At the outer disk where larger grains settle down to the midplane, mid-infrared polarized emission mainly arises from surface sub-micron grains and they are aligned with respect to the magnetic field. On the other hand, at the inner disk, micron-sized grains being aligned with the radiative flux are present at the disk surface layer. Since micron-sized grains dominate the opacity at mid-infrared, we observe the azimuthal polarization vector. As a result, we see alignment with the radiation direction at the inner disk and with the magnetic field at the outer disk. When we increase the number of magnetic inclusions, the boundary radius between the radial and the azimuthal polarization vectors decreases. This is because increasing the number of magnetic inclusions increases the maximum grain size aligned with magnetic field.
It should be noted that in this case, the disk polarization is less than 1\%, while Figure \ref{fig:ID15} shows a larger polarization degree. This is because the direction of the radiation anisotropy is perpendicular to the toroidal magnetic field; hence, the emission arising from grains aligned with the magnetic field depolarizes the emission arising from grains aligned with the direction of radiation.

In the presence of superparamagnetic inclusions ($f_p=0.1$, $\phi_{\rm sp}=0.03$), dust grains align with the magnetic field all over the disk at mid-infrared.

At millimeter wavelength, even in the presence of superparamegnetic inclusions, dust grains become aligned with the radiation direction, not with the magnetic field as one can be expected from Figure \ref{fig:timescales}.

\subsection{Grain size and wavelength dependence} \label{sec:sizedep}
\begin{figure}[t]
\begin{center}
\includegraphics[height=6.0cm,keepaspectratio]{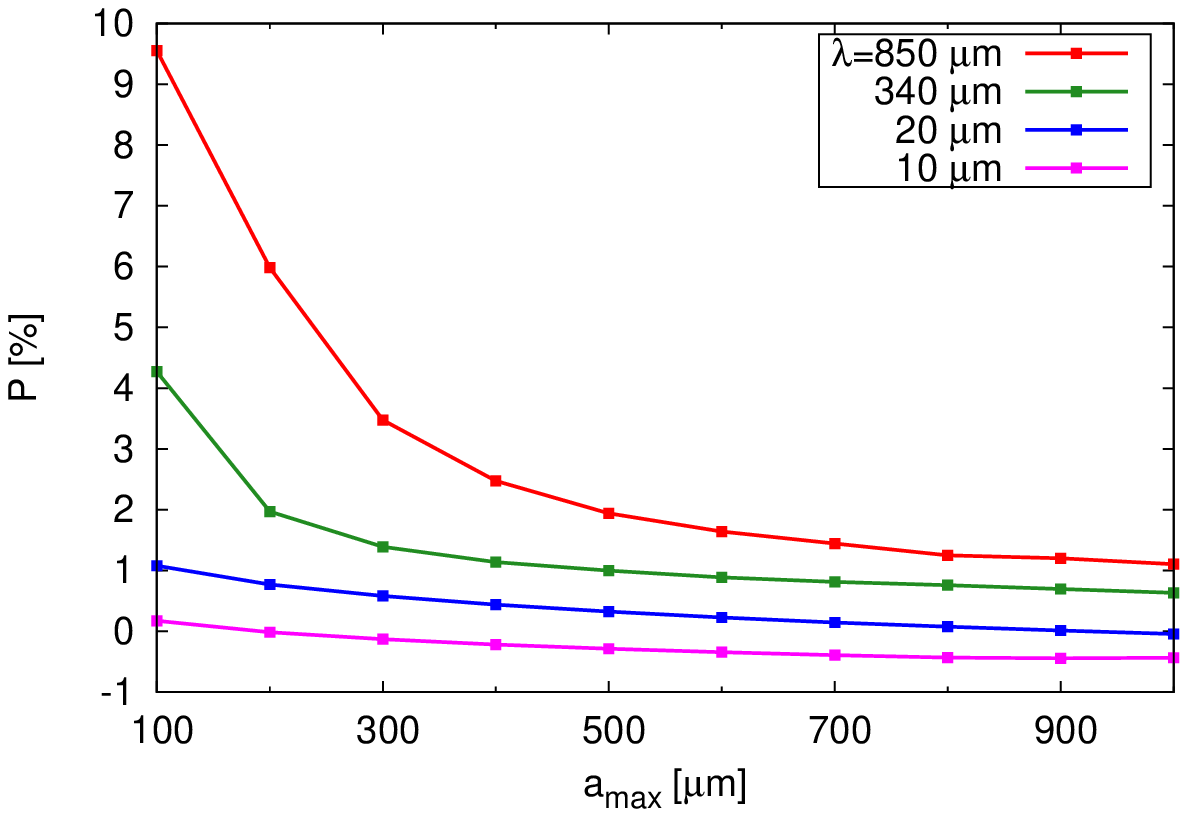}
\caption{Degree of polarization as a function of maximum grain size. Red, green, blue, and magenta lines indicate the result at $\lambda=870\ \mu$m, $\lambda=340\ \mu$m, $\lambda=20\ \mu$m, and $\lambda=10\ \mu$m, respectively. Positive and negative values represent the alignment with radiation direction and magnetic field, respectively.}
\label{fig:Pamax}
\end{center}
\end{figure}

Figure \ref{fig:Pamax} shows the degree of polarization against the maximum grain size assuming $f_{{\rm high-}J}=0.5$.
The degree of polarization is integrated over the whole radius of the disk.
At $\lambda=850\ \mu$m, with increasing the maximum grain size, the degree of polarization decreases.
This is because grains larger than the observing wavelength radiate unpolarized light; hence, with increasing the maximum grain size, more grains emit unpolarized light, and then the degree of polarization is reduced. At mid-infrared wavelength, only (sub-)micron-sized grains at the surface layer contribute to the polarized emission because the disk is optically thick. As a result, the maximum grain size does not strongly affect on the resultant polarization degree. This result implies that as the grain growth occurs, the degree of polarization can be small at all wavelengths, in particular for the (sub-)mm. 
The expected degree of polarization is much smaller than that presented in \citet{CL07}, who show a 36\% of degree of polarization for $a_{\rm max}=100\ \mu$m which drops to 8\% for $a_{\rm max}=1000\ \mu$m at $\lambda=850\mu$m. This difference is mostly due to the fact that \citet{CL07} assumed perfect internal alignment.

\begin{figure}[t]
\begin{center}
\includegraphics[height=6.0cm,keepaspectratio]{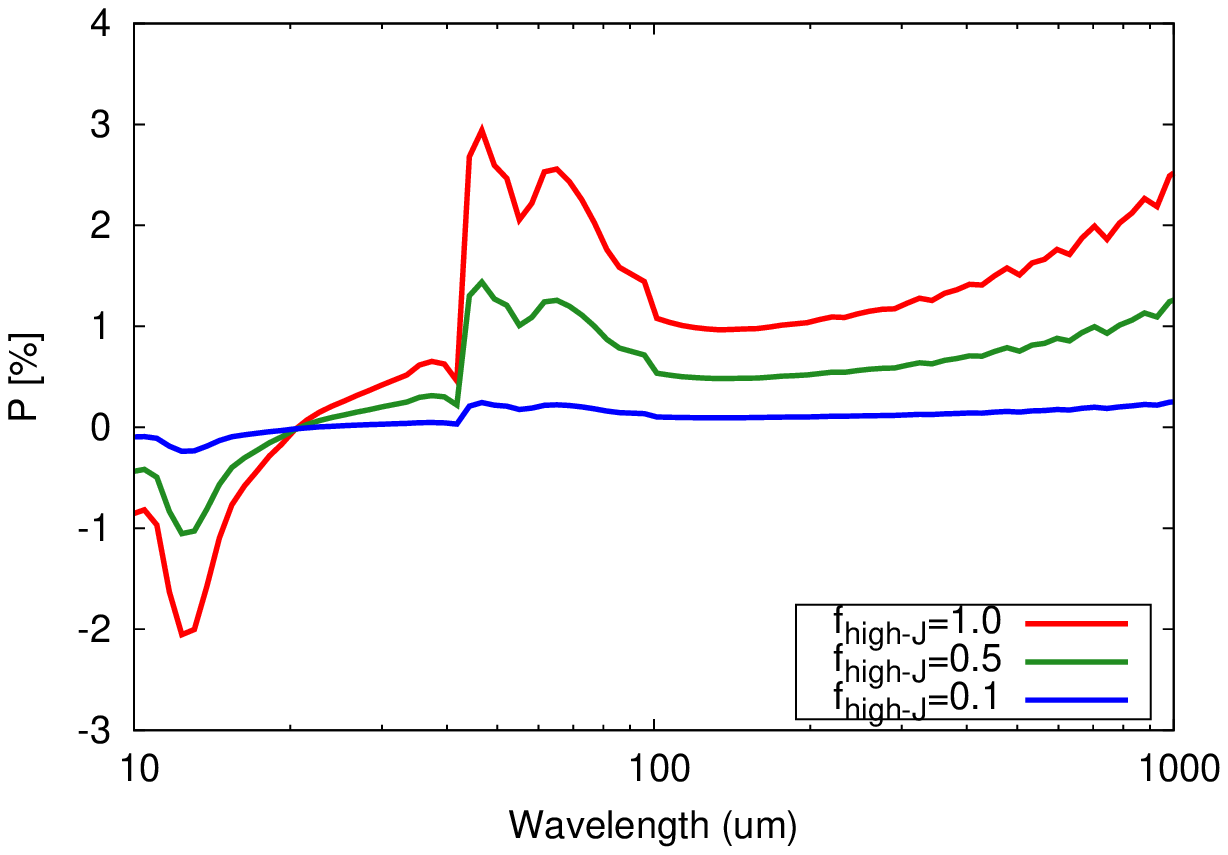}
\caption{Degree of polarization against wavelength. Red, green, and blue lines indicate $f_{\rm high-J}=1.0, 0.5,$ and $0.1$, respectively.}
\label{fig:Plambda}
\end{center}
\end{figure}

Figure \ref{fig:Plambda} shows the wavelength dependence of the degree of polarization assuming the maximum grain size to be $1000\ \mu$m.
At around $\lambda=40\ \mu$m, a strong feature appears corresponding to the ice, where refractive index changes significantly. 
At millimeter wavelength, the degree of polarization increases with increasing wavelength because more larger grains can contribute to the polarized emission for a long observing wavelength. The observed degree of polarization depends on the parameter of $f_{{\rm high-}J}$. If 90 \% of the grains become aligned with low-$J$ attractors, then the degree of polarization will be less than 1 \% at all wavelengths. This is because at low-$J$ attractors, internal alignment of the grains is poor, and then the degree of alignment is reduced significantly. 

\section{Discussion} \label{sec:discussion}

\subsection{Constraint on magnetic field strength}
As was discussed in \citet{Lazarian07} the transition from the grain alignment with respect to radiative flux to that with respect to magnetic field can be a way of determining the magnetic field (see A. Lazarian \& T. Hoang 2017, in preparation). Using this approach we can place an upper limit on the magnetic field strength if we know the radiation field, and also know that the dust is aligned with respect to the radiation. 
Using Equations (\ref{eq:tl} and \ref{eq:tradp}), the Larmor precession timescale becomes longer than the radiative precession timescale when
\begin{equation}
B \leq 59\ {\rm nG}\ a_{-5}^{\frac{3}{2}}\hat{\chi}^{-1}
\left(\frac{\bar{\lambda}}{1.2\ \mu{\rm m}}\right)\left(\frac{u_{\rm rad}}{u_{\rm ISRF}}\right)\left(\frac{\gamma \overline{|\mathbf{Q_{\Gamma}}|}}{0.01}\right) \label{eq:Bcrit}
\end{equation}
It is found that when $u_{\rm rad}=u_{\rm ISRF}$, grain alignment with radiation direction occurs when the magnetic field strength is less than 59 nG, which is much smaller than the typical value of magnetic field strength in the ISM, $\approx\mu$G. Hence, in the ISM, dust grains become aligned with the magnetic field.
At the disk surface, the energy density of the radiation is dominated by the stellar radiation; hence, substituting Equation (\ref{eq:uradsur}) into Equation (\ref{eq:Bcrit}), we obtain
\begin{equation}
B \leq 0.36 \ {\rm G}\ a_{-5}^{\frac{3}{2}}\hat{\chi}^{-1} \left(\frac{\bar{\lambda}}{1.2\ \mu{\rm m}}\right)\left(\frac{R}{10\ {\rm AU}}\right)^{-2} \left(\frac{\gamma \overline{|\mathbf{Q_{\Gamma}}|}}{0.01}\right)\label{eq:Bcrit2}
\end{equation}

We comment on the possibility of grain alignment with respect to the magnetic field at the disk midplane.
At the midplane, the wavelength of the radiation field and the grain size become much larger than the surface layer, so it is more difficult for grains to be aligned with the magnetic field. For example, when we consider the midplane at $R=50$ au of the disk, the characteristic quantities of the radiation field of our disk model are $u_{\rm rad}\simeq 1.55\times10^{-10}$ erg cm$^{-3}$, $\bar{\lambda}\simeq 137\ \mu$m, and $\gamma=1$ (see Figure \ref{fig:radfield}). Suppose the dust grains have a large number of superparamagnetic inclusions ($\phi_{\rm sp}=3$ \% and $N_{\rm cl}=10^5$), and mm-sized grains; hence we can set $Q_{\Gamma}=0.4$, and the grains can be aligned with the magnetic field when $B>B_{\rm crit}=134$ mG at $R=50$ au. The magnetic field strength adopted in this paper is $4\times 10^{-2}$ mG at $R=50$ au. Even if the toroidal magnetic field strength is amplified by 10 times, it is still too weak for grains to be aligned with respect to the magnetic field. 
It should also be stressed that, as shown in Figure \ref{fig:timescales}, the gaseous damping timescale is often shorter than the Larmor precession timescale, i.e., mm-sized grains; hence, one should also check this point whenever grain alignment with the $B$-field is assumed in disks (see Equation \ref{eq:acrit}).

\subsection{Effect of disk inclination} \label{sec:inclination}
\begin{figure}[tb]
\begin{center}
\includegraphics[height=7cm,keepaspectratio]{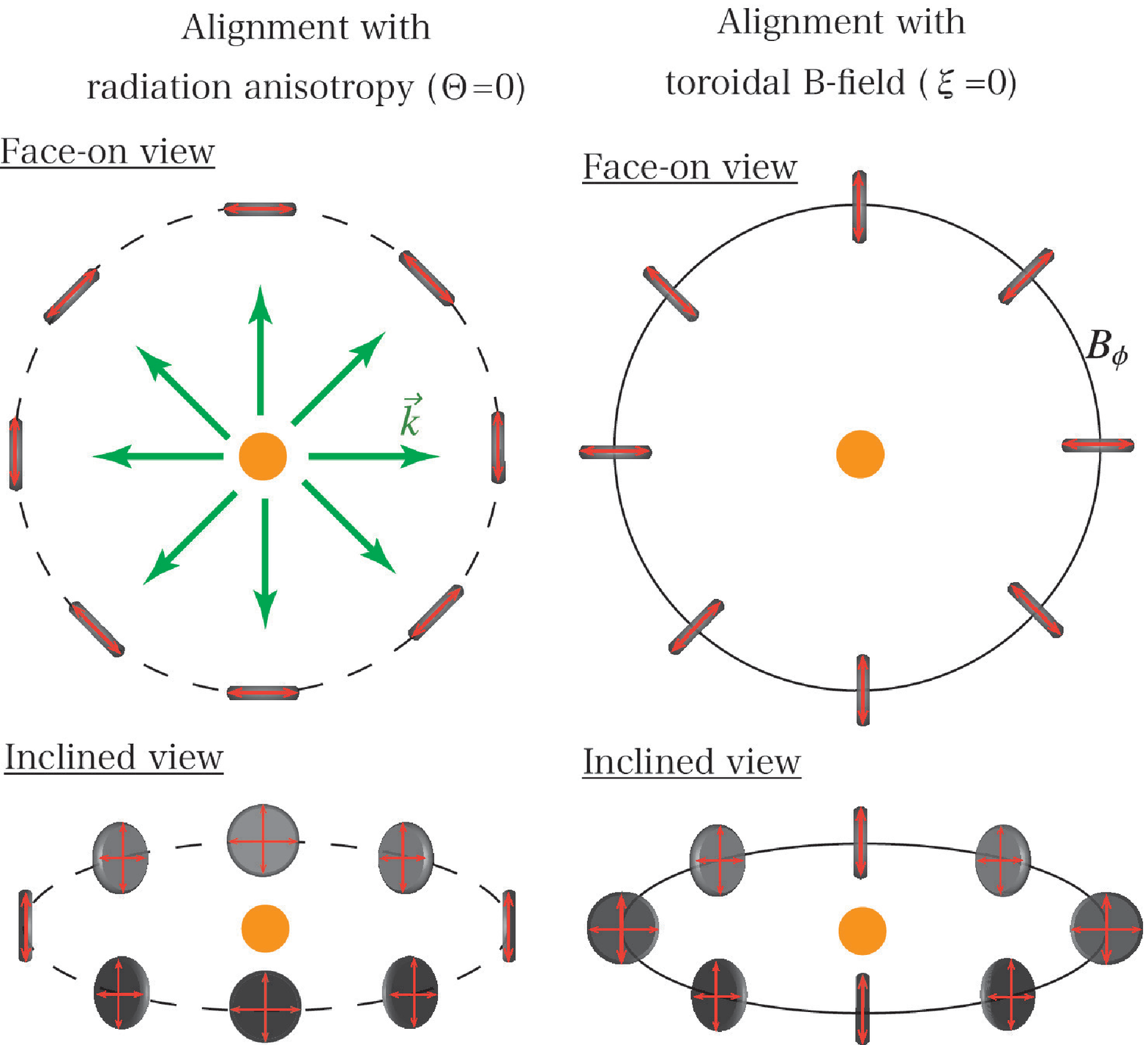}
\caption{Schematic illustration of the polarization vector arising from aligned grains. Left and right panels show the case of alignment with radiation direction and toroidal magnetic field, respectively. Top and bottom panels show face-on views and inclined view of the disk. Red arrows denote the E-vector. In this illustration, perfect internal alignment ($\theta=0$) is assumed.}
\label{fig:incl}
\end{center}
\end{figure}

In this work, we have only considered the case of face-on disks. However, most protoplanetary disks are inclined to the observer; 
hence, it is important to consider what happens for inclined disks (Figure \ref{fig:incl}).

Suppose the dust grains are aligned with the toroidal magnetic field; then the polarization vector (E-vector) would be radial for a face-on disk. For inclined disks, the disk is more polarized along its minor axis and the polarization vector is parallel to this axis (Figure \ref{fig:incl}, see also \citet{CL07}).
Along the major axis, the degree of polarization is low because the apparent shape of the oblate approaches that of a sphere.

If the dust grains align with the anisotropic radiation field, the azimuthal polarization vector can be observed for face-on disks.
For inclined disks, the disk is more polarized along its major axis, and the polarization vector would be parallel to this axis.
This tendency is qualitatively similar to the observation of HL Tau by CARMA \citep{Stephens:2014aa}.
It should be mentioned that the polarization pattern of HL Tau can also be explained by the self-scattering \citep{Kataoka:2016aa, Yang:2016aa, Yang2016b}.

\subsection{Scattering by aligned grains: circular polarization} \label{sec:V}
Circular polarization can be produced by light scattering off aligned grains \citep[see][and references therein]{Lazarian07}. 
In this case, we expect circular polarization from disks. Circular polarization occurs by (i) scattering of aligned grains, (ii) multiple scattering, or (iii) optically active chirality.
Since we are addressing emission of dust grains that are not optically active chiral, we discuss (i) and (ii) here.

The state of polarization of light is specified by four Stokes parameters, $I, Q, U, V$ \citep[e.g.,][]{Bohren:1983aa}.
Upon scattering, the Stokes parameters are changed according to the scattering matrix (or Muller matrix),
\begin{equation}
   \left(
    \begin{array}{cccc}
	I_{\rm sca} \\
	Q_{\rm sca}\\
	U_{\rm sca}\\
	V_{\rm sca}
    \end{array}
    \right)
   = \frac{1}{k^2 r^2}\left(
    \begin{array}{cccc}
      S_{11} & S_{12} & S_{13} & S_{14} \\
      S_{21} & S_{22} & S_{23} & S_{24} \\
      S_{31} & S_{32} & S_{33} & S_{34} \\
      S_{41} & S_{42} & S_{43} & S_{44}
    \end{array}
  \right)
     \left(
    \begin{array}{cccc}
	I_{\rm inc} \\
	Q_{\rm inc}\\
	U_{\rm inc}\\
	V_{\rm inc}
    \end{array} 
    \right),
\end{equation}
For aligned grains, $S_{41}$ is not zero; hence, these grains can produce circular polarization from unpolarized light upon single scattering. On the other hand, when we consider a distribution of randomly orientated grains, e.g., non-aligned grains, a symmetric treatment of the scattering matrix can reduce the non-zero independent elements, and $S_{41}=0$ for randomly orientated grains. As a result, these grains cannot circularly polarize the unpolarized light upon single scattering.
However, even in this case, since $S_{43}$ is not zero, multiple scattering can produce the circular polarization.
This is because the first scattering event produces linearly polarized light, and this light is incident on other grains.
The linearly polarized light ($Q\neq0, U=0$) for the first scattering coordinate can be an obliquely polarized light ($Q\neq0, U\neq0$) for the second scattering coordinate. As a result, $S_{43}$ can produce the circular polarization by multiple scattering. Therefore, circular polarization can be produced by either single scattering by aligned grains through non-zero $S_{41}$, or multiple scattering by randomly orientated grains through $S_{43}$. Since, at millimeter wavelength, protoplanetary disks are often optically thin, multiple scattering may not be important for many cases. 
Hence, we focus on the circular polarization due to scattering by aligned grains. 

In the Rayleigh limit (dipole approximation), the scattering matrix of an ellipsoid can be written as \citep[e.g.,][]{Bandermann:1973aa, Dolginov:1978aa, Gledhill:2000aa}
\begin{equation}
S_{41}=\frac{1}{2}ik^6(\alpha_{1}\alpha_{3}^*-\alpha_{1}^*\alpha_{3})([(\vec{e_0}\times\vec{e_1}]\cdot\hat{a_1}))(\vec{e_0}\cdot\hat{a_1}) \label{eq:S41}
\end{equation}
where $\vec{e_0}$ and $\vec{e_1}$ are the incident light and scattered light directions; hence these two vectors define the scattering plane. $\alpha$ is the polarizability of an ellipsoid with respect to each axis, given by Equation (\ref{eq:alpha}). We can see from Equation (\ref{eq:S41}) that we do not expect circular polarization for (i) non-absorbing particles (Im($\alpha)\neq0$), (ii) forward scattering and backward scattering, (iii) grains' minor axis being in the scattering plane ($[(\vec{e_0}\times\vec{e_1}]\cdot\hat{a_1})=0$), and (iv) grains' minor axis being perpendicular to the incident radiation.

Imaging observations of protoplanetary disks at visible/near-infrared wavelengths is dominated by the scattered light from disks.
Since dust grains at the surface layer are exposed to the stellar radiation, we can expect grain alignment with the radiation direction or with the magnetic field.
The degree of linear polarization can be high due to the scattering polarization, and then the polarized thermal emission is hardly detectable. However, still in this case, we can still expect the circular polarization from visible/NIR imaging.
In the solar-system, it is known that the zodiacal light is circularly polarized, and this has been explained as due to scattering by aligned grains \citep[e.g.,][]{Dolginov:1978aa, HL14}. The observation of the circular polarization in the visible/NIR may help to constrain RAT alignment in protoplanetary disks.

\subsection{Implications for Millimeter-wave Polarization}
We summarize what we can learn from the millimeter-wave polarization observations for ALMA Era.

A radially aligned polarization vector is expected for grain alignment with the toroidal magnetic field \citep{CL07, Yang:2016aa, Yang2016b, Matsakos:2016aa, Bertrang:2017aa}. 
For small grains compared to the radiation wavelength, RAT becomes inefficient, and then RAT alignment does not occur. For large grains, the Larmor precession timescale is often longer than the gaseous damping timescale. Even if Larmor precession overcomes the gaseous damping, grain alignment with the magnetic field does not occur unless the Larmor precession timescale becomes shorter than the radiative precession timescale. This suggests that the a required magnetic field strength for $B$-alignment is quite strong and dust grains should have large numbers of superparamagnetic inclusions (see Equation \ref{eq:Bcrit}). Hence, if ALMA finds the signature of the alignment with the magnetic field, this implies that the magnetic field strength in the disk might be stronger than expected so far.

If we observe a polarization vector which is consistent with alignment with the radiative flux, there are two possibilities, (i) self-scattering \citep{kataoka2015} and/or (ii) grain alignment with the radiation direction (Figure \ref{fig:ID15} or Figure \ref{fig:imagemap}). For self-scattering, we can constrain on the grain size as discussed by \citet{kataoka2015}. For grain alignment, we can constrain the upper limit on the magnetic field strength. In addition, the efficiency of RAT depends on the grain shape and size, and it might be possible to constrain grain properties. How can we distinguish between self-scattering and grain alignment observationally? As discussed in Section \ref{sec:V}, if self-scattering and grain alignment occur simultaneously, we expect circular polarization from disks. If disks do not show circularly polarized emission, disentangling between them becomes more challenging. One way to distinguish between them is to observe the degree of linear polarization.
For the self-scattering scenario, the degree of polarization is proportional to the radiation anisotropy at millimeter wavelength, hence it might be inferred from intensity observation. The degree of polarization due to grain alignment is determined by the grain axis ratio and the alignment efficiency.  As shown in Figure \ref{fig:p}, the axis ratio can increase the polarization degree. Hence, if we observe a much larger polarization degree than that estimated from the anisotropy of the radiation field, grain alignment might explain this. It is worth noting that, in principle, grain alignment is determined by the radiation anisotropy of radiation field {\it integrated over all wavelengths}, while millimeter scattering traces the anisotropy at {\it millimeter wavelength}. These two radiation fields are not necessary the same. Another solution is to perform multi-wavelength observations, because the polarized intensity due to self-scattering shows a maximum value at $\lambda_{\rm obs}\approx 2\pi{a}_{\rm max}$, and otherwise polarization intensity drops quickly to zero \citep{kataoka2015}, while grain alignment predicts weaker wavelength dependence. 

Recently, the first submillimeter polarization observation by ALMA was reported by \citet{kataoka2016b}. 
The disk shows the radial polarization vector, and the polarization vector flips by 90$^{\circ}$ at its edge.
They performed radiative transfer modeling of this object, taking self-scattering into account.
They found that the self-scattering model succeeded in reproducing the radial polarization vector as well as the flip of polarization vectors at the edge of the disk.
Since this disk has a ring-like structure, and at the center of the ring, the radiative flux can be in the azimuthal direction, and then the scattering can produce the radial polarization vector \citep[e.g.,][]{kataoka2015}. The flip of the polarization vector also be naturally explained by the change of the direction of the radiative flux. On the other hand, the self-scattering model does not reproduce the highest polarization region where the polarization degree is $13.9$ \%. According to the RAT alignment we studied in this paper, dust grains are likely to be aligned with the direction of radiative flux at (sub-)millimeter wavelength. Hence, even if we consider the RAT alignment, the observed polarization pattern of HD 142527 might be reproduced including the flip of the polarization vector. One advantage of the grain alignment model is that alignment polarization might explain the highest polarization region of HD 142527. More detailed modeling of this object taking the grain alignment into account will be left for the future study.

\section{Summary} \label{sec:summary}
We applied RAT theory of grain alignment to protoplanetary disks. Our findings can be summarized as follows.

\begin{itemize}
\item Near the star the RAT alignment is expected in the direction determined by the radiation flux rather than the magnetic field. The dust grains are expected to become aligned with their short axis to the radiation direction. If radiative alignment is taken into account, millimeter wavelength polarization can be arose from aligned grains, and the polarization vector traces the direction of radiation anisotropy.

\item Large grains in disks may not be aligned with the magnetic field as their precession rate may be less than the Larmor precession or the precession in the magnetic field. Therefore polarimetry can test the percentage of such grains.  

\item Smaller grains may align with the magnetic field at low-density regions such as the surface layer, and this might be observed in the mid-infrared polarimetry of disks. The magnetic field alignment at the surface layer occurs when dust grains have superparamagnetic inclusions or the disk magnetic field is strong  (Equation \ref{eq:Bcrit2}).

\item The degree of millimeter-wave polarization depends mainly on maximum grain size $a_{\rm max}$, axis ratio $s$, and $f_{\rm high-J}$. For $a_{\rm max}=100\ \mu$m, we expect $P\approx 10$\% polarization degree, but for $a_{\rm max}=1000\ \mu$m the polarization degree drops to $P\approx 1$\%.

\item The degree of mid-infrared polarization mainly depends on the magnetic susceptivity at zero frequency $\chi(0)$($f_{\rm p}$ and $\phi_{\rm sp}$), axis ratio $s$, and $f_{\rm high-J}$. The radial dependence of mid-infrared polarization is sensitive to the strength of turbulent.

\end{itemize}

\acknowledgments
We appreciate useful comments by the referee.
R.T. is supported by the Research Fellowship from JSPS for Young Scientists (15J02840).
A.L. acknowledges NSF grant AST 1109295, NASA grant NNH 08ZDA0090. 
This work was supported by Grants-in-Aid for Scientific Research 23103005 and 25400229, and the International collaboration program at Tokyo Tech (HN). R.T. thanks to Akimasa Kataoka, C. P. Dullemond, Hidekazu Tanaka, and Satoshi Okuzumi for useful discussions.

\bibliography{alignment}

\end{document}